 \documentclass[final,5p,times,twocolumn,authoryear]{elsarticle}

\usepackage[T1]{fontenc}
\usepackage[utf8]{inputenc}

\usepackage{graphicx}
\usepackage{amssymb}
\usepackage{amsmath}
\usepackage{textgreek}

\usepackage{multirow}
\usepackage{multicol}
\usepackage{fourier}
\usepackage{booktabs}
\usepackage{threeparttablex}
\usepackage{longtable}
\usepackage{xfrac}
\usepackage{booktabs}

\renewcommand{\cite}{\citep}

\usepackage{lineno}

\usepackage{subcaption}

\usepackage{upgreek}

\biboptions{round}

\journal{arXiv}

\begin{document}

\begin{frontmatter}

%% Title, authors and addresses

\title{Spatial sampling of MEG and EEG revisited: From spatial-frequency spectra to model-informed sampling}

%% use the tnoteref command within \title for footnotes;
%% use the tnotetext command for the associated footnote;
%% use the fnref command within \author or \address for footnotes;
%% use the fntext command for the associated footnote;
%% use the corref command within \author for corresponding author footnotes;
%% use the cortext command for the associated footnote;
%% use the ead command for the email address,
%% and the form \ead[url] for the home page:
%%
%% \title{Title\tnoteref{label1}}
%% \tnotetext[label1]{}
%% \author{Name\corref{cor1}\fnref{label2}}
%% \ead{email address}
%% \ead[url]{home page}
%% \fntext[label2]{}
%% \cortext[cor1]{}
%% \address{Address\fnref{label3}}
%% \fntext[label3]{}

%% use optional labels to link authors explicitly to addresses:
%% \author[label1,label2]{<author name>}
%% \address[label1]{<address>}
%% \address[label2]{<address>}

\author[nbe]{Joonas Iivanainen\corref{cor}}
\ead{joonas.iivanainen@aalto.fi}
\author[nbe]{Antti J.~Mäkinen\corref{cor}}
\ead{antti.makinen@aalto.fi}
\author[nbe]{Rasmus Zetter}
\author[nbe]{Matti Stenroos}
\author[nbe]{Risto J.~Ilmoniemi}
\author[nbe]{Lauri Parkkonen}

\address[nbe]{Department of Neuroscience and Biomedical Engineering, Aalto University School of Science, FI-00076 Aalto, Finland}

\cortext[cor]{These authors contributed equally to this work.}

\begin{abstract}
In this paper, we analyze spatial sampling of electro- (EEG) magnetoencephalography (MEG), where the electric or magnetic field is typically sampled on a curved surface such as the scalp. Using simulated measurements, we study the spatial-frequency content in EEG as well as in on- and off-scalp MEG. The analysis suggests that on-scalp MEG would generally benefit from three times more samples than EEG or off-scalp MEG. Based on the theory of Gaussian processes and experimental design, we suggest an approach to obtain sampling locations on surfaces that are optimal with respect to prior assumptions. Additionally, the approach allows to control, e.g., the uniformity of the sampling locations in the grid. By simulating the performance of grids constructed with different priors, we show that for a low number of spatial samples, model-informed non-uniform sampling can be beneficial. For a large number of samples, uniform sampling grids yield nearly the same total information as the model-informed grids. 

\end{abstract}

\begin{keyword}
 magnetoencephalography \sep electroencephalography \sep  on-scalp MEG  \sep spatial sampling  \sep Gaussian process \sep spatial frequency

%% keywords here, in the form: keyword \sep keyword
% We simulate the electric potential and magnetic field using a realistic head model. 
%% MSC codes here, in the form: \MSC code \sep code
%% or \MSC[2008] code \sep code (2000 is the default)

\end{keyword}

\end{frontmatter}

%% Start line numbering here if you want
%\linenumbers

%\listoftodos

\section{Introduction}
\label{sec:intro}
\noindent Sufficient sampling of a continuous signal ensures that the signal can be reconstructed from the samples without loss of relevant information. Theorems that dictate the necessary conditions for accurate signal reconstruction, such as the number of samples, are essential for guiding data acquisition (e.g., \citealt{shannon1949,pesenson2000sampling, mcewen2011novel}). The best known of these theorems is the Shannon--Nyquist sampling theorem of bandlimited functions \cite{shannon1949, luke1999, unser2000}, which connects continuous 1-D signals and their discrete representations. The theorem also applies to separable multi-dimensional signals such as two- and three-dimensional images.

The necessary conditions imposed by sampling theorems cannot always be followed in spatial sampling. For example, there can be limitations on the number of samples and their locations or the properties of the signal may not be fully known a priori. Further, if the sampling domain is not simple, there may be no applicable general theorem to guide the sampling; the optimal sampling positions have to be determined by other means, e.g., with statistical methods \cite{krause2008near}.

In this work, we analyze spatial sampling of electro- (EEG) and magnetoencephalography (MEG), where the electric and magnetic fields of the brain are measured on surfaces embedded in 3-D. We carry out a spatial-frequency analysis of continuous field patterns using the eigenfunctions of the surface Laplace--Beltrami operator \cite{reuter2009, bronstein2017geometric}. The basis formed by these functions can be seen as a natural generalization of the 1-D Fourier basis on a surface. We further describe how more compact bases can be constructed using prior information of the neuronal field patterns. We discuss how these basis-function representations of the field patterns relate to the number of spatial samples.

We investigate how to obtain optimal sample positions in EEG and MEG. We utilize Gaussian processes \cite{abrahamsen1997review, williams2006} to encode prior knowledge and to make Bayesian inference of the field. This perspective is similar to kriging in geospatial sciences \cite{chiles2018} and Gaussian-process regression in machine learning \cite{williams2006}. We introduce measures from experimental design \cite{lindley1956measure, chaloner1995bayesian, krause2008near} that can be used to quantify the optimality of the sample positions. We suggest a method that maximizes Shannon's information \cite{shannon1949} for a given number of samples and prior assumptions. We use the method to generate sampling grids for different EEG and MEG experiments, where either the whole brain or parts of it are of interest.

\section{Spatial sampling of EEG and MEG}
\label{sec:review}

\noindent EEG and MEG are noninvasive neuroimaging techniques for measuring brain activity at millisecond time scale \cite{hamalainen1993magnetoencephalography, nunez2006electric}. EEG and MEG measure the electric and magnetic field, respectively, due to neuronal currents. Adequate spatial sampling is necessary to capture the spatial detail of the continuous field due to brain activity. EEG setups typically involve 16--128 electrodes placed uniformly on the subject's scalp while state-of-the-art MEG systems use around 300 superconducting quantum interference device (SQUID) sensors placed rigidly around the subject's head \cite{baillet2017magnetoencephalography}.

Spatial sampling of EEG and MEG has again become a topic of interest. High-density EEG (hdEEG) with an electrode count of 128--256 or even more has been suggested to be beneficial (e.g., \citealt{brodbeck2011electroencephalographic,petrov2014ultra,grover2016,hedrich2017comparison,robinson2017very}). In MEG, in contrast to conventional liquid-helium-cooled SQUID sensors that measure the field $\sim$2 cm off the scalp, novel sensors such as optically-pumped magnetometers (OPMs; \citealt{budker2007optical}) and high-T$_\mathrm{c}$ SQUIDs \cite{faley2017high} have enabled on-scalp field sensing within millimetres from the head surface. Simulations have shown that the closer proximity of the sensors to the brain improves spatial resolution of MEG and increases information about neuronal currents \cite{schneiderman2014information, boto_potential_2016, iivanainen_measuring_2017, riaz2017evaluation}. The number of sensors that is beneficial in on-scalp MEG is not currently well-known. A recent study suggested that fewer sensors are needed in on-scalp MEG to achieve similar spatial discrimination performance as with SQUIDs \cite{tierney2019pragmatic}, while another study suggested with experiments that $\sim$50 OPMs give comparable results with a state-of-the-art SQUID system \cite{hill2020multi}.

The number of sensors or, equivalently, sensor spacing in EEG has been extensively studied. Srinivasan and colleagues (\citeyear{srinivasan1998}) argued that to characterize the full range of spatial detail available for EEG, at least 128 electrodes are needed (spacing $\sim$3 cm). Simulations have shown that the spatial low-pass filtering by the layered conductivity structure of the head limits useful sensor spacing in EEG \cite{srinivasan1996}. More recently, finite-element modeling in a realistic head model suggested a minimum electrode spacing of 50--59 mm determined by the distance at which the potential decreased to 10$\%$ of its peak \cite{slutzky2010optimal}. Experiments by Freeman and colleagues (\citeyear{freeman2003spatial}) suggested that electrode spacing as small as 5--8 mm could be beneficial, if high spatial frequencies contain behaviorally relevant information above the noise level.

The sensor spacing in MEG has been less studied. Analytical calculations in half-infinite homogeneous volume conductors suggest that the sensor spacing should be approximately equal to the distance of the sensors to the brain \cite{ahonen1993}. Whether volume conduction affects the number of spatial degrees of freedom in MEG in a realistic geometry is not known, as volume conduction has been studied mainly from the perspective of head model accuracy (e.g., \citealt{stenroos2014comparison}). In many MEG studies, different sensor arrays have been compared without necessarily formulating the comparison as a sampling problem \cite{wilson2007comparison,nenonen2004total,boto_potential_2016, iivanainen_measuring_2017, riaz2017evaluation}.

In EEG and MEG, previous studies have mainly focused on spatial sampling where the sensors cover the entire scalp uniformly. However, the field can also be sampled nonuniformly. Additionally, in some applications, such as brain--computer interfaces, only specific parts of the brain may be of interest. In those applications, targeted sensor arrays that extract information of local cortical activity would be of value: one could reduce the sensor count by placing sensors such that the spatial resolution and sensitivity to the cortical area of interest are maximized.

\section{Theory}

\subsection{Spatial-frequency representation of a topography}
\label{sec:sfbasis}
\noindent The quasi-static electric potential or magnetic field component is generated by distributed source activity $\vec{q}(\vec{r}_\mathrm{s})$ inside the brain. We call this field pattern on a surface a \textit{topography}; it can be written in terms of Green's function $\vec{\mathcal{L}}(\vec{r},\vec{r}_\mathrm{s})$ as \cite{hamalainen1993magnetoencephalography, nunez2006electric}
\begin{equation}
\label{eq:leadfield}
     t(\vec{r}) = \int \vec{q}(\vec{r}_\mathrm{s}) \cdot \vec{\mathcal{L}}(\vec{r},\vec{r}_\mathrm{s}) dV_\mathrm{s}.
\end{equation}
By discretizing the neural source distribution $\vec{q}(\vec{r}_\mathrm{s})$ into a set of primary current dipoles $\vec q_i(\vec r_i)=q_i \hat{q}_i \delta(\vec r - \vec r_i), i=1,\ldots,M$, where $q_i$ is the amplitude of the $i$th source at $\vec r_i$ and $\hat{q}_i$ its orientation, Eq. \eqref{eq:leadfield} becomes
\begin{equation}
\label{eq:discleadfield}
    t(\vec{r}) = \sum_i q_i \hat{q}_i \cdot \vec{\mathcal{L}}(\vec{r},\vec{r}_i ) = \sum_i q_i t_i(\vec{r}), 
\end{equation}
where $t_i(\vec{r}) = \hat{q}_i \cdot \vec{\mathcal{L}}(\vec{r},\vec{r}_i )$ is the \textit{source topography}, i.e., topography due to a unit source $\hat{q}_i$ at $\vec{r}_i$.  When the topography is also discretized to (or sampled at) $N$ points, the equation reduces to a linear matrix equation $\mathbf{t} = \mathbf{L} \mathbf{q}$, where $\mathbf{L}$ is the $(N \times M)$ lead-field matrix \cite{hamalainen1993magnetoencephalography}.

In MEG and EEG, we sample the topographies $t(\vec{r})$ on a curved 2-D surface embedded in 3-D space. Conventional 1-D Fourier analysis can be extended to spatial-frequency (SF) analysis on such surfaces using a suitable orthonormal function basis. The generating equation for the spatial-frequency basis \cite{levy2006} $\{u_m\}$ is the Helmholtz equation
\begin{equation}
    -\nabla_{\mathrm{LB}}^2 u_m = k_m^2 u_m,
    \label{eq:eigen_lb}
\end{equation}
where $\nabla_{\mathrm{LB}}^2$ is the Laplace--Beltrami (LB) operator, i.e., the surface Laplacian. The equation is an eigenvalue equation for $-\nabla_{\mathrm{LB}}^2$, where the eigenfunctions $u_m$ represent the modes of standing waves on the surface (see Fig. \ref{fig:bfs}) and eigenvalues $k_m^2$ are the squared (spatial) frequencies. These functions have been used in a variety of applications from signal and geometry processing \cite{levy2006, reuter2009} to cortical analysis \cite{qiu2006smooth} and deep learning \cite{bronstein2017geometric}. On a compact surface, the eigenvalue spectrum is discrete $\{k_m,u_m\}$ and the eigenfunctions form an orthonormal basis with respect to the inner product \cite{levy2006}:
\begin{equation}
    \langle u_i, u_j  \rangle = \int_S u_i(\vec r) u_j(\vec r) \mathrm{d}S = \delta_{ij},
    \label{eq:innerproduct}
\end{equation}
where $\delta_{ij}$ is the Kronecker delta.

The SF basis can be used to analyze the spatial-frequency content of the topographies. As the basis is orthonormal, any topography $t(\vec{r})$ can be expressed as a linear combination of SF basis functions
\begin{equation}
    t(\vec{r}) = \sum_{m=1}^\infty a_m u_m(\vec r)\, = \sum_{m=1}^\infty \langle t, u_m  \rangle u_m(\vec r)\, ,
    \label{eq:toposfrep}
\end{equation} where the coefficients are projections of $t(\vec{r})$ on $\{u_m\}$. Due to orthonormality, topography energy can be decomposed as
\begin{equation}
    ||t||^2 = \langle t, t  \rangle =  \int_S |t(\vec r)|^2 \mathrm{d}S = \sum_{m=1}^\infty a_m^2\,,
    \label{eq:parseval}
\end{equation}
a relation similar to Parseval's theorem for Fourier series. The squared coefficients $a_m^2 = \langle t, u_m  \rangle^2$ comprise the \textit{energy spectrum} of the topography in the sense of spectral signal analysis. 

If the topography $t$ is solely due to source activity (Eq. \eqref{eq:discleadfield}), the spectral coefficients can be written as $a_m = \sum_{i} q_i \langle t_i, u_m \rangle$. Assuming that the source amplitudes $q_i$ are independent random variables with zero mean ($\operatorname{E}(q_i q_j)=0$), the expected energy of $t$ can be decomposed using the spectra of the source topographies
\begin{equation}
\label{eq:expecenergytopo}
    \operatorname{E}(\|t\|^2) = \sum_{m=1}^\infty \sum_{i} \operatorname{E}(q_i^2) \langle t_i, u_m \rangle^2\,.
\end{equation}

%\begin{equation}
%    \|t\|_{L^2}^2 = \sum_m \sum_{i,j} q_i q_j \langle t_i, u_m \rangle \langle t_j, u_m \rangle\,.
%\end{equation}

\begin{figure*}[!t]
\centering
\includegraphics[width=\linewidth]{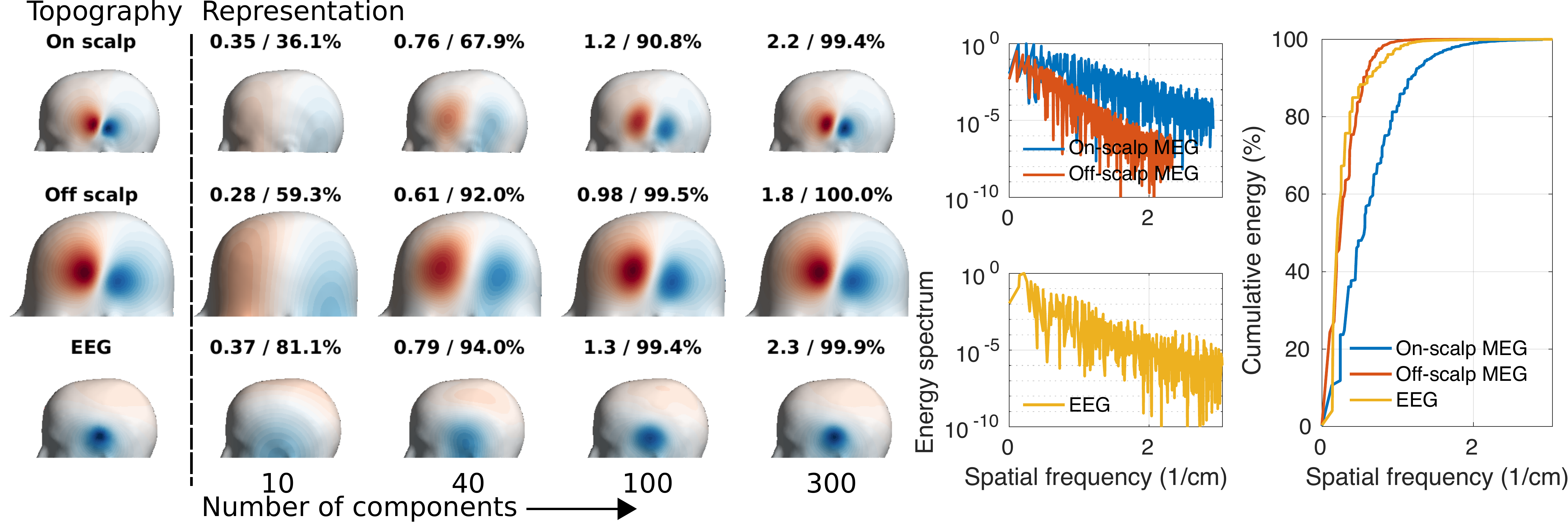}
\caption{Spatial-frequency representations of topographies in MEG and EEG. 'On scalp' refers to MEG measurement within millimetres from the scalp while 'off scalp' to that within 2 cm. For details, see Sec. \ref{sec:models_and_field_computation}. {\bf Left:} Representation of a topography with an increasing number of spatial-frequency components. The maximum spatial frequency (1/cm) and the cumulative energy of the representation are shown. {\bf Right:} Energy spectrum and cumulative energy of the topography as a function of spatial frequency.}
\label{fig:spatfrerepres}
\end{figure*}

Sampling theorems typically assume that the signals are bandlimited [e.g., in 1-D \cite{shannon1949}; on a sphere \cite{mcewen2011novel}; on a surface/manifold \cite{pesenson2014multiresolution,pesenson2015sampling}]. In our context, a bandlimited topography can be expressed with a limited number of SF basis functions $\{u_m\}$, i.e., $a_m=0$ for $m>B$ ($k_m > k_B)$. Figure \ref{fig:spatfrerepres} shows the spatial-frequency representation of a topography in MEG and EEG. Although most of the topography energy in general resides at low spatial frequencies, topographies are not strictly bandlimited. Thereby, standard sampling theorems cannot be directly applied to them.

To determine an effective bandlimit for a topography $t$, we express it as
\begin{equation}
\begin{split}
\label{eq:topoblrepres}
    t(\vec r) &= t_B(\vec{r}) + t_\mathrm{r}(\vec{r}) = \sum_{m=1}^B a_m u_m(\vec r) + \sum_{m=B+1}^\infty a_m u_m(\vec r)\,,
    %\\ &= \boldsymbol{\psi}(\vec r)^\top\mathbf{a} + \boldsymbol{\psi}_\mathrm{r}(\vec r)^\top\mathbf{a}_\mathrm{r}, 
\end{split}
\end{equation}
where $t_B$ is the $B$-bandlimited representation of the topography and $t_\mathrm{r}$ the residual. %$\|t_B\|_{L^2}^2 = \sum_{m=1}^B \langle t, u_m\rangle^2$ as the cumulative energy up to $B$. 
If $B$ is chosen such that every source topography $t_i$ is expressed up to $1-\alpha$ of its total energy (e.g., $1-0.01=0.99$), the expected energy of the residual topography (the truncation error) can be bounded using Eq. \eqref{eq:expecenergytopo} as
\begin{equation}
\begin{split}
    \operatorname{E}(\|t_\mathrm{r}\|^2) &=  \sum_{i} \operatorname{E}(q_i^2) \sum_{m=B+1}^{\infty}  \langle t_i, u_m \rangle^2 \\ &\leq \sum_{i} \operatorname{E}(q_i^2) \alpha \| t_i\|^2 = \alpha \operatorname{E}(\|t\|^2)\,.
\end{split}
\end{equation}
Thus, any topography due to independent neural sources is expected to have $1-\alpha$ of its energy in the subspace spanned by the $B$ first SF basis functions. One way to obtain an effective bandlimit $B$ (and an effective sample spacing) for bounding the expected error is to study the topography spectra of the individual sources.

\subsection{Sampling and reconstruction assuming a bandlimit}
\label{sec:sampling_finite}
\noindent 
Here, we further analyze how an effective bandlimit $B$ can be chosen by studying the reconstruction of a topography from noisy samples. We first assume that the topography can be approximated with the $B$ first SF basis functions as $t\approx t_\mathrm{B}$. To simplify the notation of Eq.~\eqref{eq:topoblrepres}, we express the summation over the basis functions as a dot product $t_B(\vec{r}) = \mathbf{u}(\vec r)^{\top}\mathbf{a}$ where $\mathbf{u}(\vec r)[m] = u_m( \vec r)$ and $\mathbf{a}[m] = a_m$.

We model $N$ noisy samples of $t_\mathrm{B}( \vec r)$ at locations $\vec r_n$ stacked in a column vector $\boldsymbol{y}$ as
\begin{equation}
    \mathbf{y}  
    = \begin{bmatrix} 
    \mathbf{u}({\vec r_1})^\top\\
    \vdots \\
    \mathbf{u}({\vec r_N})^\top \\
    \end{bmatrix}
    \mathbf{a} 
    + \boldsymbol{\epsilon}
    = \mathbf{U} \mathbf{a} + \boldsymbol{\epsilon},
    \label{eq:matrixmeas}
\end{equation}
where $\mathbf{U}$ is an $(N \times B)$ matrix with the column vectors $\mathbf{u}$ at $\vec r_n$ on rows, i.e., $\mathbf{U}[n,m] = u_m ( \vec r_n)$, and $\boldsymbol{\epsilon} \sim \mathcal{N}(0, \sigma^2\mathbf{I})$ is white measurement noise. When $N\geq B$, we can estimate the coefficients $\mathbf{a}$ using the least-squares method
\begin{equation}
    \hat{\mathbf{a}} = (\mathbf{U}^\top\mathbf{U})^{-1}\mathbf{U}^\top \mathbf{y} = \mathbf{G}\mathbf{y}\,,   \label{eq:ord_least_squares}
\end{equation}
which minimizes the sum of squared errors between the measurements and the model. The bandlimited topography $t_\mathrm{B}$ can be reconstructed using the estimated coefficients as
\begin{equation}
    \hat{t}_\mathrm{B}(\vec r) = \mathbf{u}(\vec r)^\top \hat{\mathbf{a}} =  \mathbf{u}(\vec r)^\top \mathbf{G}\mathbf{y} = \sum_n \left(\sum_m \mathbf{G}[m,n]u_m(\vec r)\right) y_n\,,
    \label{eq:ord_least_squares_interp}
\end{equation}
where the functions $\sum_m \mathbf{G}[m,n]u_m(\vec r)$ are interpolation functions for the samples $y_n$. 

If the reconstruction $\hat{t}_\mathrm{B}$ matches the continuous topography $t$ on the whole sampling surface, we can argue that sampling is adequate. To study the reconstruction error, we update the measurement model as
    $\mathbf{y} = \mathbf{U}\mathbf{a} + \mathbf{U}_\mathrm{r}\mathbf{a}_\mathrm{r} + \boldsymbol{\epsilon},$
where $\mathbf{U}_\mathrm{r}[n,m] = u_{m+B}(\vec{r}_n)$ corresponds to the residual topography $t_\mathrm{r}$. Inserting $\mathbf{y}$ into Eq.~\eqref{eq:ord_least_squares}, the coefficient estimate can be expressed as
\begin{equation}
    \hat{\mathbf{a}} = \mathbf{G}( \mathbf{U}\mathbf{a} +\mathbf{U}_\mathrm{r}\mathbf{a}_\mathrm{r} + \boldsymbol{\epsilon}) 
    =  \mathbf{a} + \mathbf{G}(\mathbf{U}_\mathrm{r}\mathbf{a}_\mathrm{r} + \boldsymbol{\epsilon}) = \mathbf{a} + \Delta\mathbf{a}\,,
\end{equation}
where $\Delta\mathbf{a} = \mathbf{G}(\mathbf{U}_\mathrm{r}\mathbf{a}_\mathrm{r} + \boldsymbol{\epsilon})$ is the error in the estimated coefficients. The error consists of two parts: $\mathbf{G}\mathbf{U}_\mathrm{r}\mathbf{a}_\mathrm{r}$ is the error due to aliasing from components outside the band $m\leq B$, and  $\mathbf{G}\boldsymbol{\epsilon}$ is the random error due to noise. 

The expected reconstruction error can now be evaluated as
\begin{equation}
\label{eq:fieldrecoerr0}
    \operatorname{E}\|t - \hat{t}_\mathrm{B}\|^2 =  \operatorname{E}\|t_B - \hat{t}_\mathrm{B}\|^2 + \operatorname{E}\|t_\mathrm{r}\|^2 =
    %\operatorname{E}\|\mathbf{a}-\hat{\mathbf{a}}\|^2 + \|\mathbf{a}_\mathrm{r}\|^2 =
    \operatorname{E}\|\Delta\mathbf{a}\|^2 + \operatorname{E}\|\mathbf{a}_\mathrm{r}\|^2\,.
\end{equation}
Since $\operatorname{E}(\|\mathbf{G}\boldsymbol{\epsilon}\|^2) = \operatorname{Tr}[\sigma^2({\mathbf{U}^{\top}\mathbf{U}})^{-1}]$ and $\operatorname{E(\boldsymbol{\epsilon})}=0$, we can rewrite the error as 
\begin{equation}
\label{eq:fieldrecoerr}
\begin{split}
\operatorname{E}\|t - \hat{t}\|^2 
    &=  \operatorname{Tr}[\sigma^2({\mathbf{U}^{\top}\mathbf{U}})^{-1}] + \operatorname{E}(\|\mathbf{G}\mathbf{U}_\mathrm{r}\mathbf{a}_\mathrm{r}\|^2) + \operatorname{}E(\|\mathbf{a}_\mathrm{r}\|^2)\,. %\\
    %&\leq (1+C)\|\mathbf{a}_\mathrm{r}\|^2 + \operatorname{Tr}[\sigma^2({\mathbf{U}^{\top}\mathbf{U}})^{-1}]\,
    \end{split}
\end{equation}
If the noise term dominates the reconstruction error, $t_\mathrm{B}$ can be considered a reasonably good approximation of $t$; the white noise variance $\sigma^2$ can then be used to determine the number of components $B$ and the number of samples $N\geq B$ needed for the reconstruction. If the noise level is unknown, a threshold such as 1\% expected residual can be used to determine the effective bandlimit $B$. As described in the previous section, this threshold may be obtained by finding the $B$ that covers at least 99\% energy of every source topography; a similar threshold has been used previously (e.g., \citealt{ahonen1993, grover2016}). Last, both the noise and the aliasing error depend on the sample positions $\{\vec{r}_n\}$ via the matrix $\mathbf{U}$. We will quantify the optimality of the sample positions in Sec. \ref{sec:optimalcriteria}.

In the presented analysis, we assumed a certain band, representing a subspace of all possible topographies, to reconstruct the topography from the samples. For a more flexible incorporation of prior assumptions, we next present a Bayesian approach to analyze the sampling and reconstruction.

\subsection{Random fields and Bayesian formulation of sampling and reconstruction}
\label{sec:bayes_sampling}

\noindent In the Bayesian framework, the prior assumptions of the signals are cast to prior probability distributions. In this section, we express the topography as $t(\vec r) = \boldsymbol{\psi}(\vec r)^\top\mathbf{a}$, where $\{\psi_m\}$ can be any set of (not necessarily orthonormal) basis functions (e.g., $u_m$ or $t_i$). We incorporate the prior knowledge by assigning a prior distribution for the coefficients $a_m$, which we consider random variables. We assume them to be Gaussian with a joint probability density $\mathcal{N}(\mathbf{m}_a, \mathbf{K}_a)$, where $\mathbf{m}_a$ is the prior mean of $\mathbf{a}$ and $\mathbf{K}_a$ is the covariance matrix of $\mathbf{a}$, representing the prior uncertainty about the coefficients.

A linear combination of basis functions with Gaussian random coefficients is a \emph{Gaussian random field}, an extension of the Gaussian process \cite{williams2006} to 3-D space. From this perspective, $t(\vec r) = \boldsymbol{\psi}(\vec r)^\top\mathbf{a}$ is a random topography, which can be described by its \emph{mean field}
  \begin{equation}
    \mu_t(\vec{r}) = \operatorname{E}(t(\vec{r})) = \sum_m    \boldsymbol{\mu}_a[m] \psi_m(\vec{r}) = \boldsymbol{\psi} (\vec{r})^\top \boldsymbol{\mu}_a
  \end{equation}
  and \textit{covariance kernel}
  \begin{equation}
    \begin{split}
      K_t(\vec{r}, \vec{r}\,') &= \operatorname{Cov}(t(\vec{r}), t(\vec{r}\,'))\\ &=\sum_{m,m'} \mathbf{K}_a[m,m']\psi_m(\vec{r}) \psi_{m'}(\vec{r}\,')  \\ &= \boldsymbol{\psi}(\vec{r})^\top \mathbf{K}_a \boldsymbol{\psi}(\vec{r}\,')
      \label{eq:kernel_def}\,.
    \end{split}
  \end{equation}
The mean field represents the expected value of the topography based on prior knowledge and the variance $K_t(\vec{r}, \vec{r}\,'=\vec{r}\,)$ describes the uncertainty around the mean. Every finite collection of samples of a random field is jointly distributed as $\mathcal{N}(\boldsymbol{\mu}, \mathbf{K})$, where $\mathbf{K}$ is the \emph{sample covariance matrix} defined elementwise as $\mathbf{K}[n,n'] = K_t(\vec{r}_n, \vec{r}_{n'}) =  \boldsymbol{\psi}(\vec{r}_n)^\top \mathbf{K}_a \boldsymbol{\psi}(\vec{r}_{n'})$, and $\boldsymbol{\mu}$ is the sample mean, i.e., $\boldsymbol{\mu}[n] = \mu_t(\vec{r}_n) = \boldsymbol{\psi} (\vec{r}_n)^\top \boldsymbol{\mu}_a$.

\begin{figure}[!t]
\centering
\includegraphics[width=1.0\linewidth]{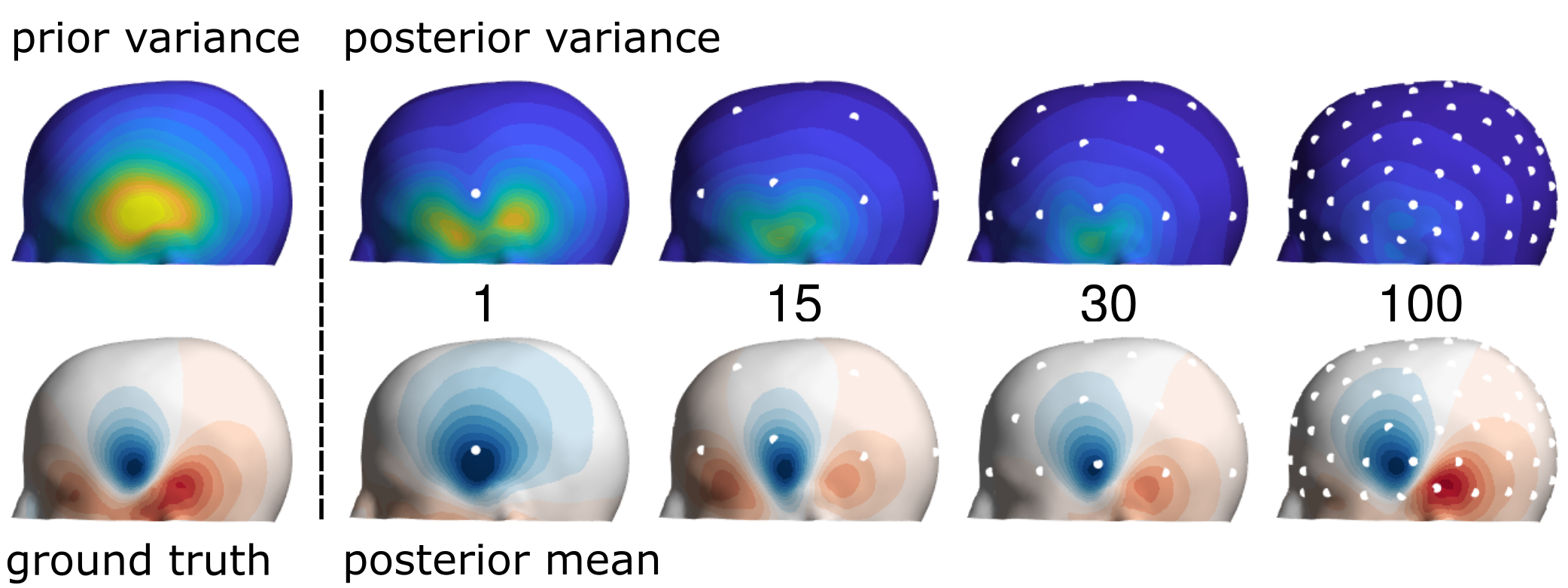}
\caption{Bayesian reconstruction of a topography. {\bf Left}: Prior variance and a ground-truth topography. {\bf Right}: Posterior variance and the topography reconstruction (the posterior mean field) after 1, 15, 30 and 100 noisy measurements (white dots) taken uniformly on the surface. Each additional sample decreases the posterior variance, yielding a better reconstruction of the topography.}
\label{fig:random_field_sampling}
\end{figure}

The effect of measurement is encoded in the posterior distribution of the topography as illustrated in Fig.~\ref{fig:random_field_sampling}. Having a collection of (noisy) samples $\mathbf{y}$ at sampling points $R = \{\vec{r}_n\,|\,n=1...N\}$, modeling the noise as  $\boldsymbol{\epsilon} \sim \mathcal{N}(0, \boldsymbol{\Sigma})$, and assuming $\operatorname{E}(t(\vec{r}))=0$, the posterior mean field and covariance are \cite{williams2006}
\begin{equation}
\begin{split}
    \mu_t(\vec{r}\,|\,\mathbf{y}, R) 
    =& \mathbf{k}(\vec{r})^\top(\mathbf{K} + \boldsymbol{\Sigma})^{-1}\mathbf{y} ,  \\
    K_t(\vec{r},  \vec{r}\,'\,|\, R) 
    =& K_t(\vec{r}, \vec{r}\,') - \mathbf{k}(\vec{r})^\top(\mathbf{K} + \boldsymbol{\Sigma})^{-1}\mathbf{k}(\vec{r}\,'),
    \label{eq:posterior_bayes}
\end{split}
\end{equation}
where $\mathbf{k}(\vec{r})$ is the covariance between the topography at the sample points and the topography at the point $\vec{r}$: $\mathbf{k}(\vec{r})[n] = K_t(\vec{r}, \vec{r}_n) = \boldsymbol{\psi}(\vec{r})^\top \mathbf{K}_a \boldsymbol{\psi}(\vec{r}_n)$. 

The posterior mean $\mu_t(\vec{r}\,|\,\mathbf{y}, R)$ depends linearly on the samples $\mathbf{y}$ and can be considered as the reconstruction of the topography similar to Eq.~\eqref{eq:ord_least_squares_interp}.
The posterior covariance $K_t(\vec{r},  \vec{r}\,'\,|\, R)$ describes the uncertainty in the topography after the measurements; the noisier the samples, the more uncertainty is left. For $\vec{r}=\vec{r}\,'$, the term subtracted from $K_t(\vec{r}, \vec{r})$ is always positive ($(\mathbf{K} + \boldsymbol{\Sigma})^{-1}$ is positive-definite). Hence, a new measurement always reduces the uncertainty about the topography (Fig.~\ref{fig:random_field_sampling}). Uncertainty is reduced at all points $\vec{r}$ that are correlated with the topography at the sampling location.

In \ref{sec:appendix_mne}, we derive the following formulas for the mean and covariance of $t(\vec r$)
\begin{equation}
\begin{split}
    \mu_t(\vec{r}\,|\,\mathbf{y}, R) 
     &= \boldsymbol{\psi}(\vec{r})^\top(\boldsymbol{\Psi^\top \Sigma^{-1} \Psi} + \mathbf{K}_a^{-1})^{-1}\boldsymbol{\Psi}^\top\boldsymbol{\Sigma}^{-1}\mathbf{y}  = \boldsymbol{\psi}(\vec{r})^\top \mathbf{\hat{a}},\\
    K_t(\vec{r},  \vec{r}\,'\,|\, R)
    &= \boldsymbol{\psi}(\vec{r})^{\top}( \boldsymbol{\Psi}^\top\boldsymbol{\Sigma}^{-1}\boldsymbol{\Psi}+ \mathbf{K}_a^{-1})^{-1} \boldsymbol{\psi}(\vec{r})\,,
    \label{eq:posteriorcoefs_bayes}
\end{split}
\end{equation}
where $\mathbf{\hat{a}} = (\boldsymbol{\Psi^\top \Sigma^{-1} \Psi} + \mathbf{K}_a^{-1})^{-1}\boldsymbol{\Psi}^\top\boldsymbol{\Sigma}^{-1}\mathbf{y}$ is the posterior mean of the coefficients and $(\boldsymbol{\Psi}^\top\boldsymbol{\Sigma}^{-1}\boldsymbol{\Psi}+ \mathbf{K}_{a}^{-1})^{-1}$ is the associated posterior covariance matrix. In the limit of white noise ($\boldsymbol{\Sigma}\rightarrow \sigma^2\mathbf{I}$) and infinite SNR ($\sigma^2 \mathbf{K}_a^{-1}\rightarrow 0$), the posterior mean reduces to the classical interpolation formula of Eq.~\eqref{eq:ord_least_squares_interp} while the posterior coefficient covariance approaches $\sigma^2(\boldsymbol{\Psi}^\top\boldsymbol{\Psi})^{-1}$, which is the covariance of the coefficient error. At this limit, the equations are well-defined only when the number of samples $N$ is larger than the number of basis functions, i.e., the topography belongs to a subspace of $B\leq N$ basis functions.

\subsection{Kernels for bioelectric potential and magnetic field}
\label{sec:bioemsampling}
\noindent In this section, we outline how to construct prior covariance kernels (Eq.~\eqref{eq:kernel_def}) for a random topography $t(\vec{r})$ on the measurement surface. In the later sections, we discuss how these kernels can be used to obtain sampling grids.

\paragraph{Spatial-frequency kernel}
The SF basis functions can be used to express the prior kernel as $K(\vec{r}, \vec{r}\,')  = \mathbf{u}(\vec{r})^\top \mathbf{K_{\mathrm{SF}}} \mathbf{u}(\vec{r}\,')$, where $\mathbf{u}(\vec{r})[i] = u_i(\vec{r})$ is the $i^\mathrm{th}$ eigenfunction of the LB operator and $\mathbf{K_{\mathrm{SF}}}$ is the covariance matrix of the spatial-frequency coefficients. To construct this kernel, knowledge of the sampling-domain geometry, i.e., of the measurement surface, is needed. For a detailed treatment of kernel representation with Laplacian eigenfunctions, we refer to the work by Solin and S\"arkk\"a (\citeyear{solin2020hilbert}).

\begin{figure*}[tbp]
\centering
\includegraphics[width=\linewidth]{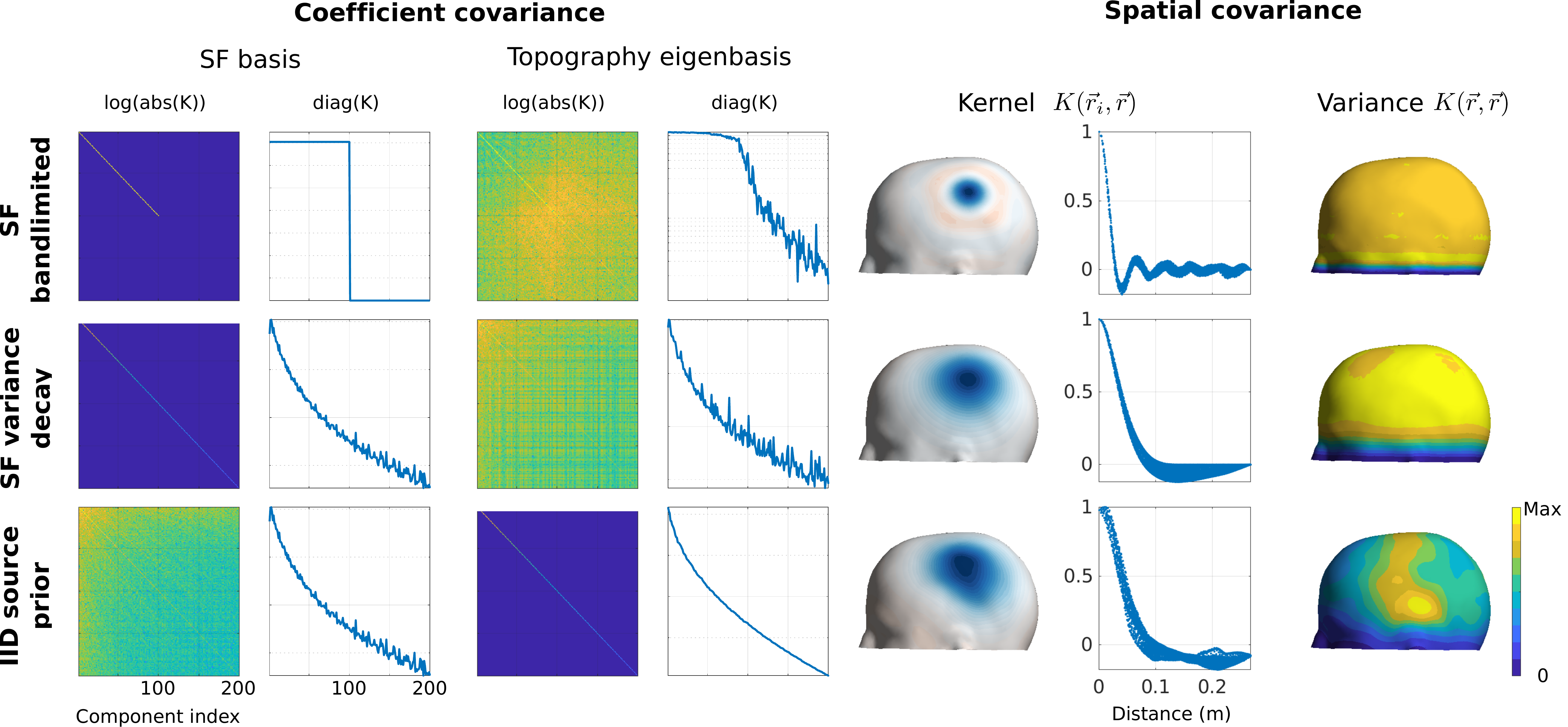}
\caption{Covariances of three priors (SF bandlimited: bandlimited in spatial-frequency basis; SF variance decay: variance roll-off in SF basis; IID source prior: identically and independently distributed neural sources generating the field). {\bf Left}: Prior coefficient covariance in the SF basis and in the topography eigenbasis of IID sources. The covariance matrix and its diagonal are shown. {\bf Right}: Spatial covariance. The spatial kernel and its decay are shown for one position on the measurement surface. The rightmost column shows the spatial variance. For details on the computation, see Sec. \ref{sec:models_and_field_computation}. Units in the plots are arbitrary.}
\label{fig:kernel}
\end{figure*}

The assumption of smoothly-varying bandlimited topographies (Sec. \ref{sec:sfbasis}) can be encoded in this form by setting the diagonal of $\mathbf{K}_\mathrm{SF}$ to a constant up to $B$ components and to zero above $B$; the highest spatial frequency of the topographies is assumed to be $k_B$. The covariance structure of this SF-bandlimited prior is visualized in Fig. \ref{fig:kernel} both in the coefficient $\mathbf{K}_\mathrm{SF}$ and spatial domains $K(\vec{r},\vec{r}\,')$. This prior for the coefficients results in covariance kernels with sinc-like spatial profiles $K(\vec{r}, \vec{r}_i)$ and uniform spatial variance $K(\vec{r}, \vec{r})$.

A physically more plausible prior kernel can be constructed by assigning the diagonal with a variance that decays towards the high spatial frequencies (SF variance decay; Fig. \ref{fig:kernel}). This results in a nearly uniform spatial variance and spatially symmetric covariance profiles without sinc-like ringing.

\paragraph{Topography kernel}
Considering the source amplitudes $q_i$ in Eq. \eqref{eq:discleadfield} to be Gaussian random variables \cite{demunck1992random}, the covariance kernel can be written as
\begin{equation}
\label{eq:fieldsourcecovar}
    K(\vec{r},\vec{r}\,') = \mathbf{t}(\vec{r})^\top \mathbf{K}_q \mathbf{t}(\vec{r}\,') = \sum_{i,j}  \mathbf{K}_q[i,j] t_i(\vec{r})t_{j}(\vec{r}\,'),
\end{equation}
where $\mathbf{K}_q$ is the $(M\times M)$ source covariance matrix. The source topographies $t_i(\vec{r})$ can be considered as the basis functions that assemble the kernel and $q_i$ as the associated (possibly correlated) random coefficients. Compared to the SF kernel, more detailed prior knowledge of the random topography can be encoded in this form. For a set of $N$ sampling points, the kernel reduces to a covariance matrix $\mathbf{K} = \mathbf{L}\mathbf{K}_q\mathbf{L}^\top$.

A relevant topography kernel can be constructed by assuming the source amplitudes $q_i$ identically and independently distributed (IID source prior): $\mathbf{K}_q = q^2\mathbf{I}$. This kernel $K_\mathrm{IID}(\vec{r},\vec{r}\,') = q^2 \boldsymbol{t}(\vec{r})^\top \boldsymbol{t}(\vec{r}\,')$ has non-uniform spatial variance and the spatial profiles $K(r,\vec{r}_i)$ are asymmetric (Fig. \ref{fig:kernel}).

\paragraph{Kernel eigendecomposition and topography eigenbasis}
Principal component analysis can be extended to random processes using the Karhunen--Loève decomposition, in which the process is expressed as a linear combination of orthonormal functions multiplied by uncorrelated random coefficients \cite{loeve1978probability, stark1986probability}; the orthonormal basis functions are obtained as the eigenfunctions of the corresponding covariance kernel \cite{mercer1909functions}. The Karhunen--Loève theorem (\citealt{stark1986probability}, section 7.6) states that this basis has the minimal truncation error among all possible bases for representing the random process with a given number of components.

To analyze MEG and EEG topographies, we investigate the eigenbasis of $K_\mathrm{IID}$. The kernel can be written using the eigendecomposition as 
\begin{equation}
\label{eq:mercer}
K_\mathrm{IID}(\vec{r},\vec{r}\,') =  \mathbf{v}(\vec{r})^\top\mathbf{D}\mathbf{v}(\vec{r}\,') = \sum_m d_m^2 v_m(\vec{r})v_m(\vec{r}\,'),    
\end{equation}
where $v_m(\vec{r})$ form the orthonormal basis on the measurement surface, which we call the \textit{topography eigenbasis}, and $\mathbf{D}$ is a diagonal matrix with variances $d_m^2$ of the eigenfunctions on the diagonal. The coefficient covariances of the different priors in the topography eigenbasis are shown in Fig. \ref{fig:kernel}. 

We can express the expected energy (or total variance) of a random topography following the IID-source assumption using the variances of the eigencomponents as $ \operatorname{E}(\|t\|^2)  = \sum_{m=1}^\infty d_m^2$. With similar arguments as in Secs. \ref{sec:sfbasis} and \ref{sec:sampling_finite}, we can get an estimate for the number of samples needed to reconstruct the random topography, e.g., by truncating the series up to the number of components $\{v_m\}$ that capture 99\% of the total variance.

\begin{figure}[t]
\centering
\includegraphics[width=\linewidth]{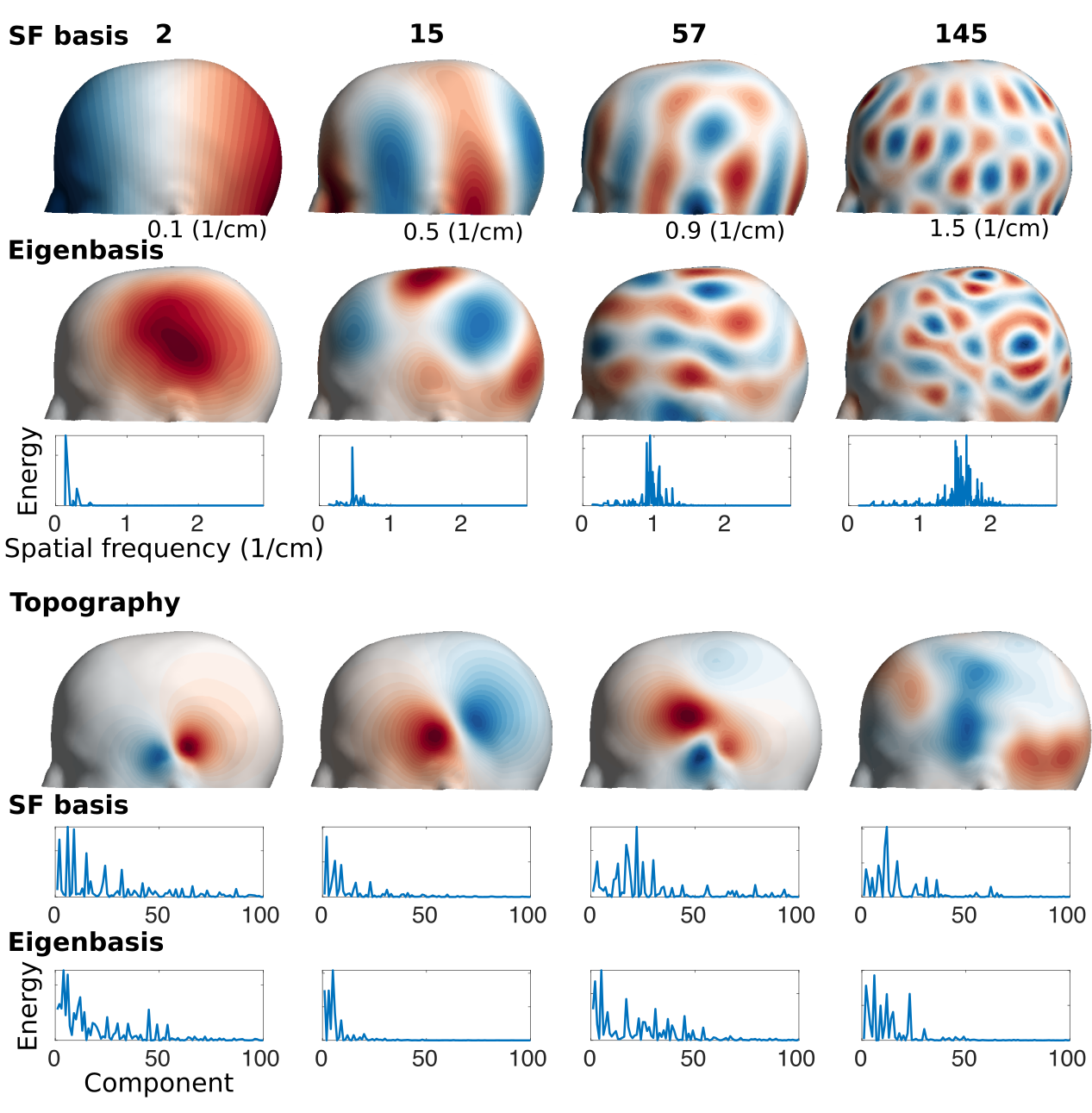}
\caption{Illustration of the orthonormal basis functions that are used to represent topographies. {\bf Top}: Representation of the spatial-frequency (SF) basis and the topography eigenbasis due to IID sources. Each column shows the SF function and eigenfunction with the corresponding index. Spatial frequencies of the SF functions are displayed. The energy spectrum of the eigenfunction in SF basis is shown at bottom. The SF basis was generated using the zero-Neumann boundary condition (Sec. \ref{sec:models_and_field_computation}). The eigenfunctions comprise multiple SF functions with an average spatial frequency increasing together with the index.  {\bf Bottom}: Random topographies and their energy spectra both in the SF basis and eigenbasis. Topography energy is distributed to lower index components in the eigenbasis than in the SF basis indicating compression by the eigenbasis.} 
\label{fig:bfs}
\end{figure}

The source topographies $t_i$ (Eq. \ref{eq:discleadfield}) can also be analyzed using the eigenbasis ($\{v_m\}$). Based on the Karhunen--Loeve theorem, we expect that $\{v_m\}$ compresses the source topography representation, e.g., on average, fewer components is needed when truncating $t_i$ in $\{v_m\}$ than in the SF basis $\{u_m\}$. Fig.~\ref{fig:bfs} gives an illustration of the two bases. We study the decomposition of source topographies in these bases in Sec.~\ref{sec:spatfreqres}.

\paragraph{Noise kernels and SNR}
Measurement noise can be modeled as a random topography $\nu(\vec{r})$ with an associated covariance kernel $K_\nu(\vec{r}, \vec{r}\,')$. Noise sources include the sensors themselves. Sensor noise is typically independent across the sensors with an equal variance $\sigma^2$, which can be modeled as an equivalent spatial white noise, i.e., a random topography with a kernel $K_\nu(\vec{r}, \vec{r}\,') = \sigma^2 \delta(\vec{r} - \vec{r}\,')$. In any orthonormal basis, the covariance of white noise is proportional to the identity matrix 
\begin{equation}
\begin{split}
    \operatorname{E}({a_i a_j}) 
    &=  \int\int u_i(\vec{r}) \sigma^ 2\delta(\vec{r} - \vec{r}\,') u_j(\vec{r}\,') dS' dS \\
    &= \sigma^2 \langle u_i, u_j \rangle = \sigma^2 \delta_{ij}.
\end{split}
\end{equation}
%&= \int\int u_i(\vec{r}) K_\nu(\vec{r}, \vec{r}\,') u_j(\vec{r}\,') dS' dS  \\ 
Noise that originates from sources other than the sensors is generally colored (i.e., not white) and can be modeled with a covariance function $K_\nu(\vec{r}, \vec{r}\,') = \boldsymbol{\psi}_\nu(\vec{r})^\top \mathbf{K}_\nu \boldsymbol{\psi}_\nu(\vec{r}\,')$, where $\boldsymbol{\psi}_\nu(\vec{r})$ are the basis functions for the noise topography and $\mathbf{K}_\nu$ is the prior covariance of the noise coefficients.

We can now define the signal-to-noise ratio (SNR) of a random topography $t(\vec{r})$. We start by diagonalizing the noise kernel as in Eq.~\eqref{eq:mercer}: $K_\nu(\vec{r}, \vec{r}\,') =  \mathbf{w}(\vec{r})^\top\boldsymbol{\Lambda}\mathbf{w}(\vec{r}\,')$, where $w_i(\vec{r})$ are eigenfunctions of $K_\nu$ and $\boldsymbol{\Lambda}$ contains the associated noise variances on its diagonal. The noise eigenfunctions can be used to compose a \textit{whitening kernel} $W(\vec{r}, \vec{r}\,') = \mathbf{w}(\vec{r})^\top\boldsymbol{\Lambda}^{-1/2}\mathbf{w}(\vec{r}\,')$, which, when applied to the noise topography $\int W(\vec{r}, \vec{r}\,') \nu(\vec{r}\,')dS'$, yields spatial white noise. Applying the whitener to a random topography, we get a whitened topography $\tilde{t}(\vec{r}) = \int W(\vec{r}, \vec{r}\,') t(\vec{r}\,')dS'$. The spatial SNR, i.e., the expected SNR for a sample at $\vec{r}$, is the variance of the whitened topography
\begin{equation}
\begin{split}
\label{eq:SNR}
\mathrm{SNR}(\vec{r})&=K_{\tilde{t}}(\vec{r}, \vec{r}) = \int\int  W(\vec{r}, \vec{r}\,') K_{t}(\vec{r}', \vec{r}'')  W(\vec{r}, \vec{r}\,'') dS' dS''.   
\end{split}
\end{equation}
With white noise, the whitening operation only acts as scaling with $1/\sigma^2$. With colored noise, it also modifies the covariance structure of the random topography.

\subsection{Optimality criteria of sampling grids}
\label{sec:optimalcriteria}
\noindent In this and the following section, we discuss how an optimal sampling grid $R$ can be constructed. A classical approach would be to choose $R$ that minimizes the expected reconstruction error, involving the inversion of $\boldsymbol{\Psi}^\top\boldsymbol{\Psi}$ (Sec. \ref{sec:sampling_finite}). In the Bayesian approach, $R$ should minimize the posterior variance, which involves the inversion of $\mathbf{K} + \boldsymbol{\Sigma}$ (Sec. \ref{sec:bayes_sampling}). One way to obtain an optimal sampling is then to find an $R$ that maximizes the "size" of either of these matrices. The matrix size can be generally quantified in multiple ways, giving different optimality criteria in the context of \textit{(Bayesian) experimental design} \cite{krause2008near, chaloner1995bayesian}.

A common criterion is \textit{A-optimality} \cite{krause2008near} measured either as $\operatorname{Tr}[(\boldsymbol{\Psi}^\top\boldsymbol{\Psi})^{-1}]$ or more generally as $\operatorname{Tr}[(\boldsymbol{\Psi}^\top\boldsymbol{\Sigma}^{-1}\boldsymbol{\Psi}+ \mathbf{K}_{a}^{-1})^{-1}]$. However, when studying sampling on a bounded surface as in MEG and EEG, we should rather measure how well the sampling pattern captures the overall variance on the surface. To optimize the measurement in this sense, we define the fractional explained variance:
\begin{equation}
    \mathrm{FEV}(R) = 1- \frac{\int K_f(\vec{r},  \vec{r}\,|\, R) dS}{\int K_f(\vec{r}, \vec{r}) dS} 
    = \frac{\int \mathbf{k}(\vec{r})^\top(\mathbf{K} + \boldsymbol{\Sigma})^{-1}\mathbf{k}(\vec{r})dS}{\int K_f(\vec{r}, \vec{r}) dS} ,
    \label{eq:fev}
\end{equation}
which ranges from 0 to 1, i.e., from no to all variance explained. This measure is related to \textit{I-optimality} or integrated optimality \cite{atkinson2014optimal}.

The matrix size can also be measured using the determinant; this criterion is called \textit{D-optimality} \cite{chaloner1995bayesian}. In the Bayesian setting, this leads to maximization of the total information (\ref{sec:apppendix_info})
\begin{equation}
    \mathrm{TI}(R) = \frac{1}{2} \log_2 \frac{\det(\mathbf{K} + \boldsymbol{\Sigma})}{\det(\boldsymbol{\Sigma})} = \frac{1}{2} \log_2 \det(\tilde{\mathbf{K}} + \mathbf{I}),
    \label{eq:info}
\end{equation}
where $\tilde{\mathbf{K}} = \boldsymbol{\Sigma}^{-1/2} \mathbf{K} \boldsymbol{\Sigma}^{-1/2}$ is the whitened sample covariance matrix. The D-optimal $R$ minimizes the posterior entropy of the random-field coefficients $\mathbf{a}$ \cite{sebastiani2000maximum}, i.e., maximizes the information gained, e.g., from the neural sources. Previously, total information has been used in studies comparing MEG sensor arrays \cite{kemppainen1989, nenonen2004total,schneiderman2014information,iivanainen_measuring_2017, riaz2017evaluation}.

\subsection{Sampling grid construction}
\label{sec:sampling_grid_construction}

\noindent In this section, we suggest a method to obtain sampling grids that maximize the total information for given prior assumptions. In \ref{sec:apppendix_info}, we show that this corresponds to maximizing the diagonal elements in the whitened sample covariance matrix $\tilde{\mathbf{K}}$ while simultaneously minimizing the absolute values of the non-diagonal elements. In other words, maximizing total information is equivalent to finding the sampling grid with the least correlations and maximal SNR. 

\begin{figure}
    \centering
    \includegraphics[width=\linewidth]{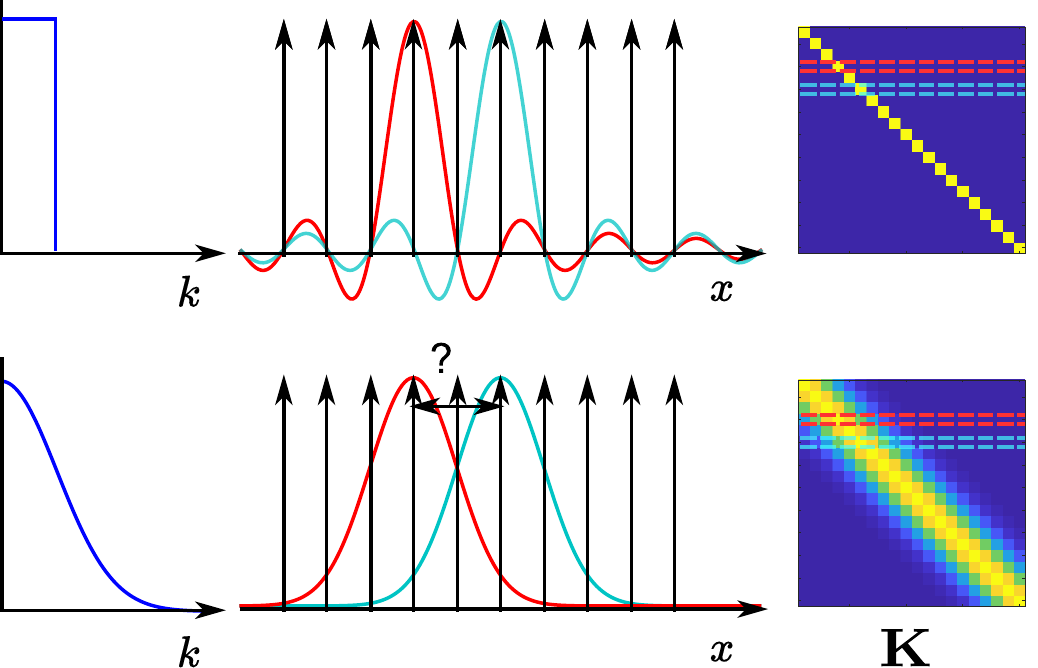}
    \caption{Illustration of covariance kernel and sampling. {\bf Left}: Coefficient variance as a function of spatial frequency for a strictly bandlimited process and a process with a smooth variance decay. {\bf Middle}: The covariance functions and samples. The red and cyan curves describe the covariance functions for two samples while the spikes show the sample positions. Equispaced samples fit the zeros of the sinc functions. For a non-sinc covariance, there is no trivial choice for the optimal sampling distance. {\bf Right}: Sample covariance matrices for the two cases, i.e., the covariance functions sampled at the equispaced locations. Dashed lines indicate the rows of the two samples.}
    \label{fig:sinc_gaussian}
\end{figure}

As an illustrating example, we first discuss how the sample spacing in the Shannon--Nyquist theorem \cite{shannon1949,jerri1977shannon} can be seen to result from information maximization. When treating bandlimited functions as random processes with a uniform prior variance for the spatial-frequency coefficients up to $k_B$, the kernel $K(x, x')$ can be calculated as the Fourier transformation of a boxcar function. This results in a sinc-function kernel with zeros at equispaced intervals $1/(2k_\mathrm{B})$ illustrated in Fig.~\ref{fig:sinc_gaussian}. When the samples are placed at the zero crossings, the sample covariance matrix $\mathbf{K}$ becomes diagonal, maximizing the total information.

To maximize the total information for a given number of samples with a more general covariance kernel, we reformulate the problem using the kernel eigenfunctions $v_n(\vec{r})$. We consider a (whitened) kernel $K(\vec{r}, \vec{r}\,')$ with an eigendecomposition $\mathbf{v}(\vec{r})^\top\mathbf{D}\mathbf{v}(\vec{r}\,')$. The decomposition can be rewritten as $K(\vec{r}, \vec{r}\,') = \mathbf{\tilde{v}}(\vec{r})^\top\mathbf{\tilde{v}}(\vec{r}\,')$, where $\mathbf{\tilde{v}}(\vec{r}) = \mathbf{D}^{1/2}\mathbf{v}(\vec{r})$, i.e., the covariance between points $\vec{r}$ and $\vec{r}\,'$ can be calculated as an Euclidean dot product between $\mathbf{\tilde{v}}(\vec{r})$ and $\mathbf{\tilde{v}}(\vec{r}\,')$. This is similar to the kernel trick used in machine learning \cite{williams2006}, where $\mathbf{\tilde{v}}(\vec{r})$ are called feature vectors. Rewriting the squared distance between these vectors as
\begin{equation}
\begin{split}
    ||\mathbf{\tilde{v}}(\vec{r}) - \mathbf{\tilde{v}}(\vec{r}\,')||^2 &= ||\mathbf{\hat{v}}(\vec{r})||^2 + || \mathbf{\tilde{v}}(\vec{r}\,')||^2 - 2\mathbf{\tilde{v}}(\vec{r})^\top\mathbf{\hat{v}}(\vec{r}\,')\\ &= K(\vec{r}, \vec{r}) + K(\vec{r}\,', \vec{r}\,') - 2K(\vec{r}, \vec{r}\,'),
\end{split}
\end{equation}
we can see that maximizing this distance with respect $\vec{r}$ and $\vec{r}\,'$ corresponds to maximizing the diagonal elements in the sample covariance matrix $\mathbf{K}$ while simultaneously minimizing the non-diagonal elements. 

The sample configuration that yields maximal information can be found with the \emph{farthest-point sampling} algorithm \cite{eldar1997farthest, schlomer2011farthest}, which attempts to maximize the pair-wise minimum distances between the sample points. Instead of maximizing distances $|\vec{r}-\vec{r}\,'|$, we maximize $\|\mathbf{\tilde{v}}(\vec{r}) - \mathbf{\tilde{v}}(\vec{r}\,')\|$; otherwise the algorithm works similarly. If $\boldsymbol{K}[i,j]=K(\vec{r}_i,\vec{r}_j)$ are positive for neighbouring samples, the algorithm leads to minimization of the non-diagonal $|\boldsymbol{K}[i,j]|$, and thereby to maximization of the total information.

Generally, the sampling grid given by the method follows the covariance structure of the kernel (Fig. \ref{fig:kernel}). For example, approximately uniform sampling grids can be generated using spatial-frequency-bandlimited covariance with circularly symmetric sinc-like kernels (Sec.~\ref{sec:methods_sampling_algorithm}). This showcases a connection between our sampling method and the sampling theorems that assume bandlimited functions and use uniform sampling \cite{shannon1949, pesenson2015sampling}.

\section{Simulations}

\noindent In this Section, we simulate spatial-frequency (SF) spectra of source topographies and covariance kernels of random topographies due to random source distributions. Further, we construct optimal sampling grids for different kernels and evaluate their performance. We begin by outlining the numerical methods.

\subsection{Models and computations}
\label{sec:models_and_field_computation}

\noindent We used a head model built using the example data of SimNIBS software (version 2.0; \citealt{windhoff2013electric}) in an earlier study \cite{stenroos2016}. The head model consists of the outer surfaces of the white matter in left and right hemispheres, gray matter, cerebellum, scalp as well as the inner and outer surfaces of the skull.

We assumed a piecewise constant isotropic conductor and computed the electric potential and magnetic field using a linear Galerkin boundary-element method with the isolated-source approach \cite{stenroos2007matlab,stenroos2012bioelectromagnetic}. We used two source spaces discretized into point-like dipolar source-current elements. To analyze the spatial frequency content of the source topographies independent of the source orientation, we constructed a volumetric source space comprising sets of three orthogonal dipoles at 4-mm spacing inside the gray-matter surface (10 635 positions). For all other analyses, we used a cortically-constrained source space with dipoles distributed on the boundary of white and gray matter at 3-mm average spacing. The 20 324 sources were oriented normal to the surface following the anatomical orientation of apical dendrites of pyramidal neurons.

We constructed three volume conductor models from the anatomical head model (see Fig. \ref{fig:spatfreqsources}C). The model 4C used in most simulations had the following compartments: brain (inside the outer surface of gray matter) and cerebellum, cerebrospinal fluid (CSF), skull and scalp. The conductivity of the soft tissues (brain, cerebellum and scalp) was set to 0.33 S/m while the conductivity of CSF was 1.79 S/m. For skull conductivity, we used the value 0.33/50 S/m. Two other volume conductor models  were constructed by removing compartments from 4C in order to illustrate how volume conduction affects the source topographies. In a three-compartment model (3C), CSF was removed from 4C so that the conductivity boundaries were defined by the inner and outer skull surfaces as well as the scalp surface (brain, skull and scalp). In a single-compartment conductor (1C), all other compartments were removed expect the scalp surface.

We calculated the electric potential and the normal component of the magnetic field at the nodes of triangular meshes that represented the measurement surfaces. The measurement surfaces were generated from a dense scalp mesh by cutting the mesh above a plane defined roughly by the ears and nose. The cut mesh was resampled to 5 404 nodes and 10 421 triangles. The mesh for the electric potential (EEG) was obtained by projecting the node positions on the scalp. An 'on-scalp' MEG surface was generated by inflating the mesh 4.5 mm away from the scalp. A more distant 'off-scalp' MEG surface was obtained by further inflating the mesh and by smoothing it with the function \texttt{smoothsurf} in the iso2mesh MATLAB toolbox \cite{fang2009tetrahedral}. The median distance of the nodes of the off-scalp MEG surface to the scalp was 2.4 cm (2.0--4.4 cm).

The SF basis vectors were computed by discretizing the LB operator to the triangle mesh in the weak form \cite{reuter2009}. The discrete form of the eigenvalue equation \eqref{eq:eigen_lb} is
\begin{equation}
\label{eq:disclb}
    -\mathbf{C}\mathbf{u}_i = k^2_i \mathbf{M u}_i,
\end{equation}
where $\mathbf{u}_i$ contains the nodal values for the $i^\mathrm{th}$ SF function, $\mathbf{M}$ is a matrix that takes account the overlap in the piecewise-linear basis functions, and $\mathbf{C}$ is the discrete LB operator. Matrices $\mathbf{C}$ and $\mathbf{M}$ were computed using MATLAB functions \texttt{cotmatrix} and \texttt{massmatrix} included in gptoolbox \cite{gptoolbox}. The zero-Neumann boundary condition, which sets the outwards-facing derivative of $u$ to zero, was used in the spectral energy analysis, while the zero-Dirichlet boundary condition, setting $u$ to zero, was used in the grid construction. The inner product of Eq.~\eqref{eq:innerproduct} discretizes to
\begin{equation}
    \langle u_i, u_j  \rangle = \int_S u_i(\vec r) u_j(\vec r) \mathrm{d}S \approx \mathbf{u}_i^\top \mathbf{M} \mathbf{u}_j.
\end{equation}

The IID source topography kernel ($K_\mathrm{IID}$; Sec.~\ref{sec:bioemsampling}) was constructed using the cortically-constrained source space and 4C. The eigenbasis of $K_\mathrm{IID}$ was computed by discretizing the kernel on the surface and solving the discrete eigenvectors $\mathbf{v}_i$.

\subsection{Energy spectra and component numbers of source topographies}
\label{sec:spatfreqres}

\begin{figure*}[!ht]
\centering
\includegraphics[width=0.75\linewidth]{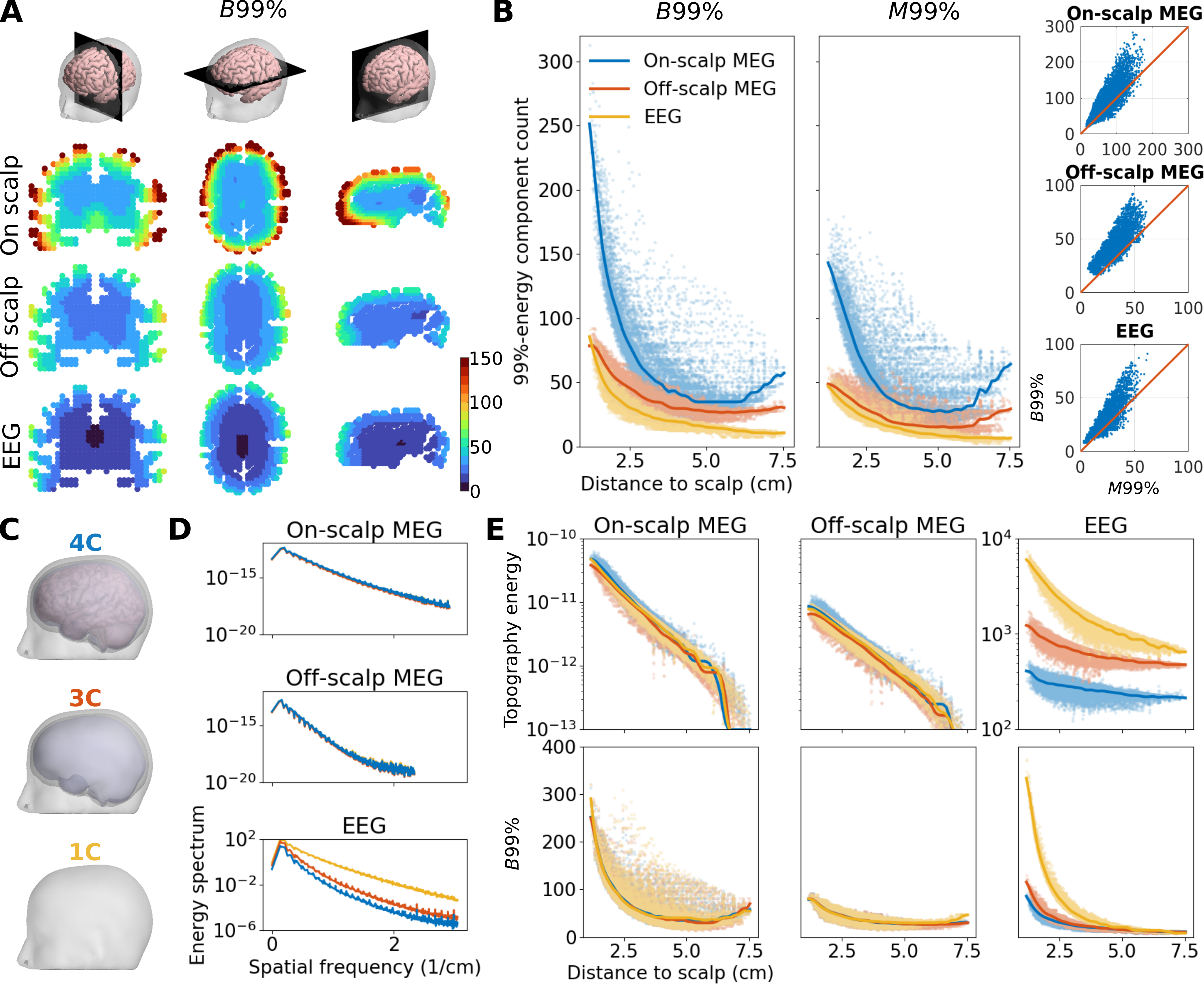}
\caption{Basis-function representations of topographies of volumetric sources (A--B) and their dependency on the volume conductor (C--E). {\bf A:} Number of spatial-frequency (SF) components to reach 99$\%$ of the energy ($B_{99\%}$) for each source position visualised on orthogonal slices of brain volume. {\bf B:} 99$\%$-energy component count in SF ($B_{99\%}$) and eigenbasis ($M_{99\%}$) as a function of source depth. The dots correspond to individual sources and the lines estimate the medians of the distributions as a function of source depth. Scatterplots across the source positions compare $M_{99\%}$ and $B_{99\%}$. {\bf C:} Volume conductors with different number of compartments (4C, 3C and 1C). {\bf D:} Energy spectra of the topographies averaged across the sources with different conductor models. {\bf E:} Topography energy and $B_{99\%}$ with different conductor models.}
\label{fig:spatfreqsources}
\end{figure*}

\paragraph{Methods} 
\noindent We quantified the source topography energies and their spectra to estimate the beneficial numbers of spatial samples as discussed in Sec. \ref{sec:sampling_finite}. The spectra were computed using the SF basis and the eigenbasis of $K_\mathrm{IID}$. Specifically, we analyzed the number of SF components needed to get 99$\%$ energy of each source topography, we denote this number by $B_{99\%}$. The maximum spatial frequency of those SF components was defined as the 99\% bandwidth of the source topography, $k_{99\%}$. In order to quantify source topography compression, we calculated the number of eigencomponents to achieve 99\% energy $M_{99\%}$ and the ratio $M_{99\%}/B_{99\%}$.

For the locations of volumetric sources, the topography energies and $B_{99\%}$ were computed by summing the squared coefficients of the three orthogonal dipoles in the same position. To see how the properties of the volume conductor affect the spectra, the analysis was repeated in all volume conductor models. To model realistic source orientations, the energy and $B_{99\%}$-analysis was repeated in the cortically-constrained source space. The sensor spacing in each modality that corresponds to the maximum $B_{99\%}$ across the sources was estimated using uniform sampling grids (see Sec. \ref{sec:methods_sampling_algorithm}).

We further inspected how many SF components were above a certain spatial white noise level. We matched the noise levels of MEG and EEG: $B_{99\%}$ components of the highest-energy source topography were above the noise level. For on-scalp and off-scalp MEG, the same noise level was used.

\paragraph{Results}
Fig. \ref{fig:spatfreqsources}A--B displays $B_{99\%}$ for each volumetric source position. Generally, $B_{99\%}$ decreases as a function of source depth. For the most superficial sources, $B_{99\%}$ ranges from 200 to 280 in on-scalp MEG, while in off-scalp MEG and in EEG it is up to 90. The bandwidth $k_{99\%}$ of the most superficial sources in on-scalp MEG is about 2.1 1/cm, while in off-scalp MEG and EEG it is approximately 1.0 and 1.2 1/cm, respectively.

\begin{figure*}[!ht]
\centering
\includegraphics[width=0.75\linewidth]{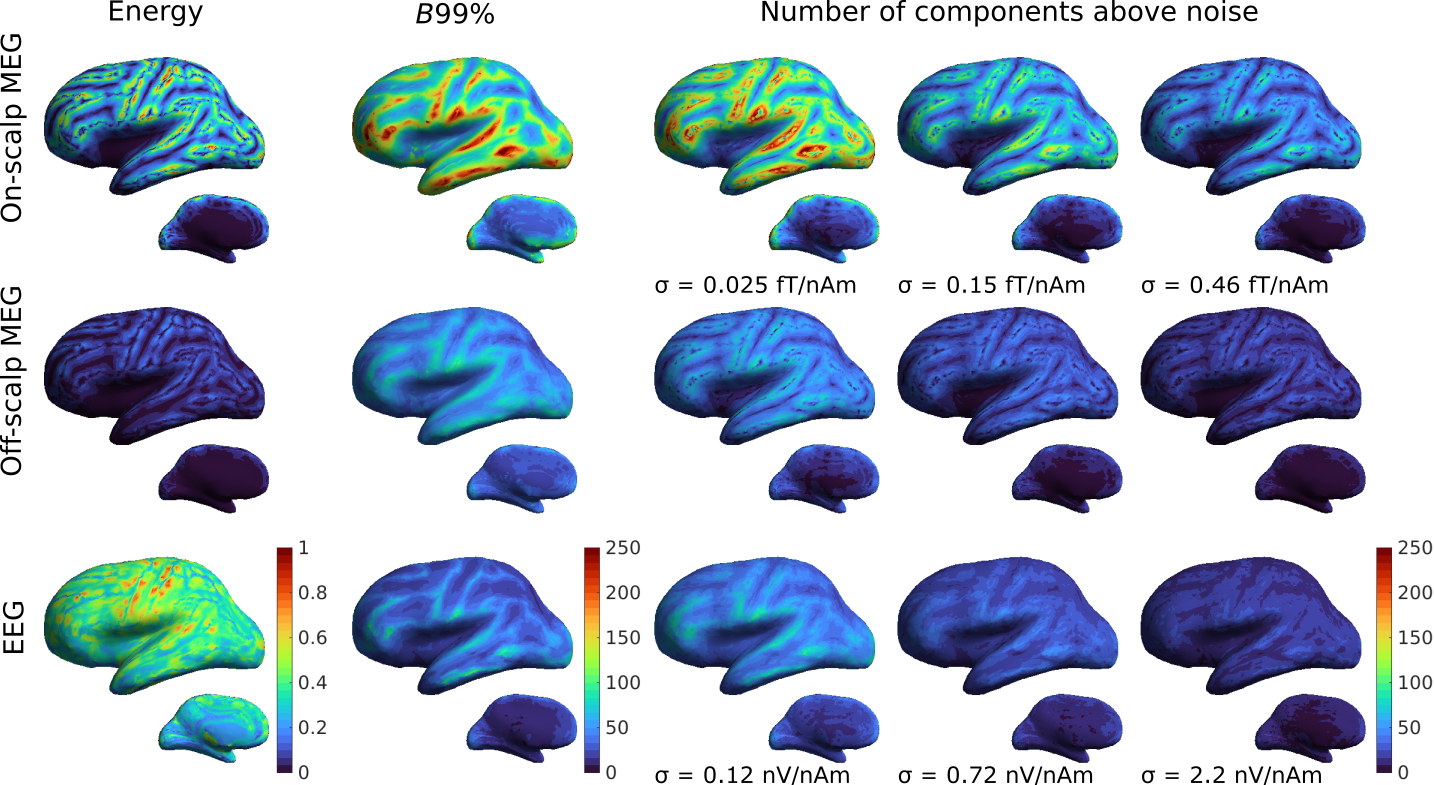}
\caption{Topography energy and spatial-frequency component count of cortically-constrained sources. {\bf Left:} Normalized topography energy (same normalization for on-scalp and off-scalp MEG). {\bf Center:} Number of components needed to capture 99\% topography energy ($B_{99\%}$). {\bf Right:} Number of components above the indicated spatial white noise level (given as a ratio to the source amplitude).}
\label{fig:cortcomps}
\end{figure*}

The numbers of topography eigencomponents $M_{99\%}$ are shown in Fig. \ref{fig:spatfreqsources}B. For on-scalp MEG, up to 179 components are required, while corresponding counts for the off-scalp MEG and EEG are 61 and 62. The number of  eigencomponents $M_{99\%}$ is typically lower than the number of SF components $B_{99\%}$. The ratio $M_{99\%}/B_{99\%}$ is on average 81$\%$, 67$\%$ and 73$\%$ in on-scalp MEG, off-scalp MEG and EEG with ranges 42--200$\%$, 30--130$\%$ and 30--180$\%$, respectively.

Topography energy as a function of source depth is shown in Fig. \ref{fig:spatfreqsources}E. For superficial sources, the energy is 6--7 times higher in on-scalp than in off-scalp MEG. The effect of different compartments of the volume conductor on the topography energy spectra is visualized in Fig. \ref{fig:spatfreqsources}D--E. In on-scalp and off-scalp MEG, the topography energy spectra are relatively unaffected by the CSF and skull in the volume conductor. On the contrary, topography spectra in EEG show a dependency on the spatial filtering occurring due to volume conduction: for example, in 1C, the maximum $B_{99\%}$ is roughly 350 giving $k_{99\%}$ of about 2.5 1/cm, while including the poorly-conducting skull (3C) causes spatial low-pass filtering, leading to a maximum $B_{99\%}$ of 120.

Topography energies and SF component counts for the cortically-constrained sources are shown in Fig. \ref{fig:cortcomps}. In MEG, superficial tangential sources have the highest energies. The maximum $B_{99\%}$ is about 300, 100 and 110 in on-scalp MEG, off-scalp MEG and EEG ($k_{99\%}$: 2.2, 1.0 and 1.3 1/cm). For these maximum $B_{99\%}$, the estimated uniform sample spacings are 1.5, 3.4 and 2.6 cm. With the set noise level (0.5 fT and 2.4 nV assuming a source amplitude of 20 nAm), at maximum 305, 90 and 105 SF components are above noise in on-scalp, off-scalp MEG and EEG, respectively; with an about 18 times higher noise level (9.2 fT and 44 nV) those counts are 114, 43 and 40.

\subsection{Topography kernels}

\paragraph{Methods} We analyzed the spatial variance $K_\mathrm{IID}(\vec{r}, \vec{r})$ (amount of signal) and correlations $K_\mathrm{IID}(\vec{r}_i, \vec{r})$ (sample spacing; Sec. \ref{sec:bioemsampling}). The number of eigencomponents explaining 99\% of the total variance was calculated to estimate the spatial degrees of freedom of the random source distribution (Sec. \ref{sec:bioemsampling}). Topography correlation length was quantified for each measurement point $\vec{r}_i$ by computing the distance along the surface at which $K_\mathrm{IID}(\vec{r}_i, \vec{r})$ had decayed to half of its maximum value (the 'half-maximum width'). The distances along the surfaces were computed using a method based on the heat equation \cite{Crane2017HMD} implemented as the MATLAB function \texttt{heat\_geodesic} in gptoolbox \cite{gptoolbox}.

\begin{figure}[!t]
\centering
\includegraphics[width=\linewidth]{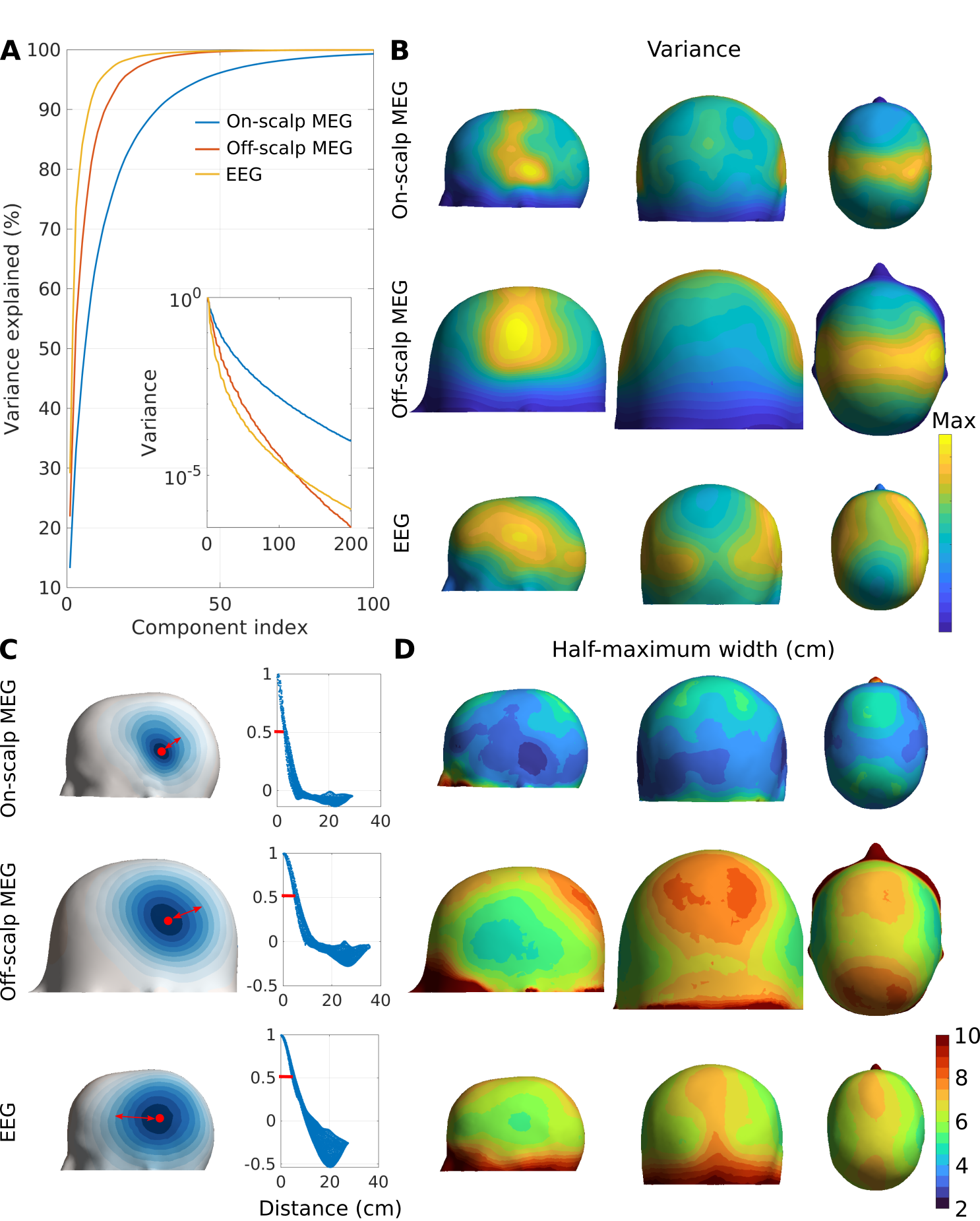}
\caption{Topography kernel due to independent Gaussian sources with an equal variance. {\bf A:} Variance as a function of the kernel eigencomponent. Inset shows the normalized eigenvalues (variances). {\bf B:} Distribution of the variance on the measurement surface. {\bf C:} Spatial profile of the field kernel at one position on the measurement surface and its decay as a function of distance along the surface. The length at which the kernel has decayed to half (half-maximum width) is illustrated. {\bf D:} Half-maximum width of the kernel across the measurement surface.}
\label{fig:fieldcovariance}
\end{figure}

\paragraph{Results} Fig. \ref{fig:fieldcovariance} illustrates the kernels of the IID random source distribution for the three modalities. To explain 99$\%$ of the total variance, 88, 35 and 25 eigencomponents are needed in on-scalp MEG, off-scalp MEG and EEG, respectively. Variance is distributed nonuniformly on the measurement surface with the highest values around the temporal cortex. The kernels are asymmetric and the correlation length varies across the surface; the least correlated area can be found on top of the temporal cortex. The correlation lengths are shortest in on-scalp MEG as indicated by smaller values of half-maximum widths compared to off-scalp MEG and EEG.

\subsection{Sampling grid construction and evaluation}

\label{sec:methods_sampling_algorithm}

\paragraph{Method} 
The sampling grids were constructed by subsampling the nodal positions of the triangle meshes using the method described in Sec.~\ref{sec:sampling_grid_construction}. For a given kernel discretized on the mesh, we calculated its eigendecomposition, and utilized a farthest-point sampling algorithm implemented in gptoolbox \cite{gptoolbox} to obtain the optimal sampling positions. We ran the optimization algorithm several times with a random initial configuration to seek the global optimum as the output of the algorithm was sensitive to the initial condition.
%With $N$ samples, we chose the number of components in the eigenspace to be $2N$.

\paragraph{Grid construction} We constructed sampling grids for scenarios where random source distributions were defined in the brain: a global scenario where the whole brain was assumed active and of interest, and for a local scenario where a region of interest (ROI) in the brain was defined. We compared the performance of uniform sampling to model-informed sampling for different numbers of spatial samples $N$. Uniform sampling was generated using bandlimited priors in the SF basis (white covariance for the $N$ lowest SF coefficients; SF bandlimited prior; Sec. \ref{sec:bioemsampling}). Model-informed sampling was generated by encoding the prior in the covariance of the cortically-constrained sources (IID source prior; Sec. \ref{sec:bioemsampling}). The grids were evaluated by computing total information (TI) and fractional explained variance FEV according to Eqs.~\eqref{eq:info} and \eqref{eq:fev} from the 'ground-truth' model of the signal and noise.

\subsubsection{Global scenario}

\paragraph{Methods} Here, the ground-truth model corresponded to the IID source prior and spatial white noise. The grids were constructed on meshes with the regions of eyes cropped out. We performed two analyses. In the first, we compared uniform and model-informed sampling in EEG and MEG as a function of SNR. The standard deviation of the white noise ranged from $0.05\sigma_0$ to $20\sigma_0$. The reference noise level $\sigma_0$ was fixed in each modality so that they gave an average SNR \eqref{eq:SNR} of 1 for source standard deviation of 7.7 pAm. Using this convention, $\sigma_0$ was determined to be 9 fT, 3.7 fT and 42 nV in on-scalp, off-scalp MEG and EEG, respectively.

In the second analysis, we compared uniform sampling of on-scalp and off-scalp MEG as a function of spatial white noise level. We set the reference source variance $q_0^2$ so that the average SNR of off-scalp MEG was 1 with a white noise level of 3 $\mathrm{fT}$. The white noise level of on-scalp MEG ranged from 1 to 20 $\mathrm{fT}$ and the source variance was varied ($0.2q_0^2$, $q_0^2$ and $5 q_0^2$). On-scalp MEG grids comprising 10--600 points were generated, while off-scalp MEG had 100 or 300 points.

\begin{figure*}[!t]
\centering
\includegraphics[width=1.0\linewidth]{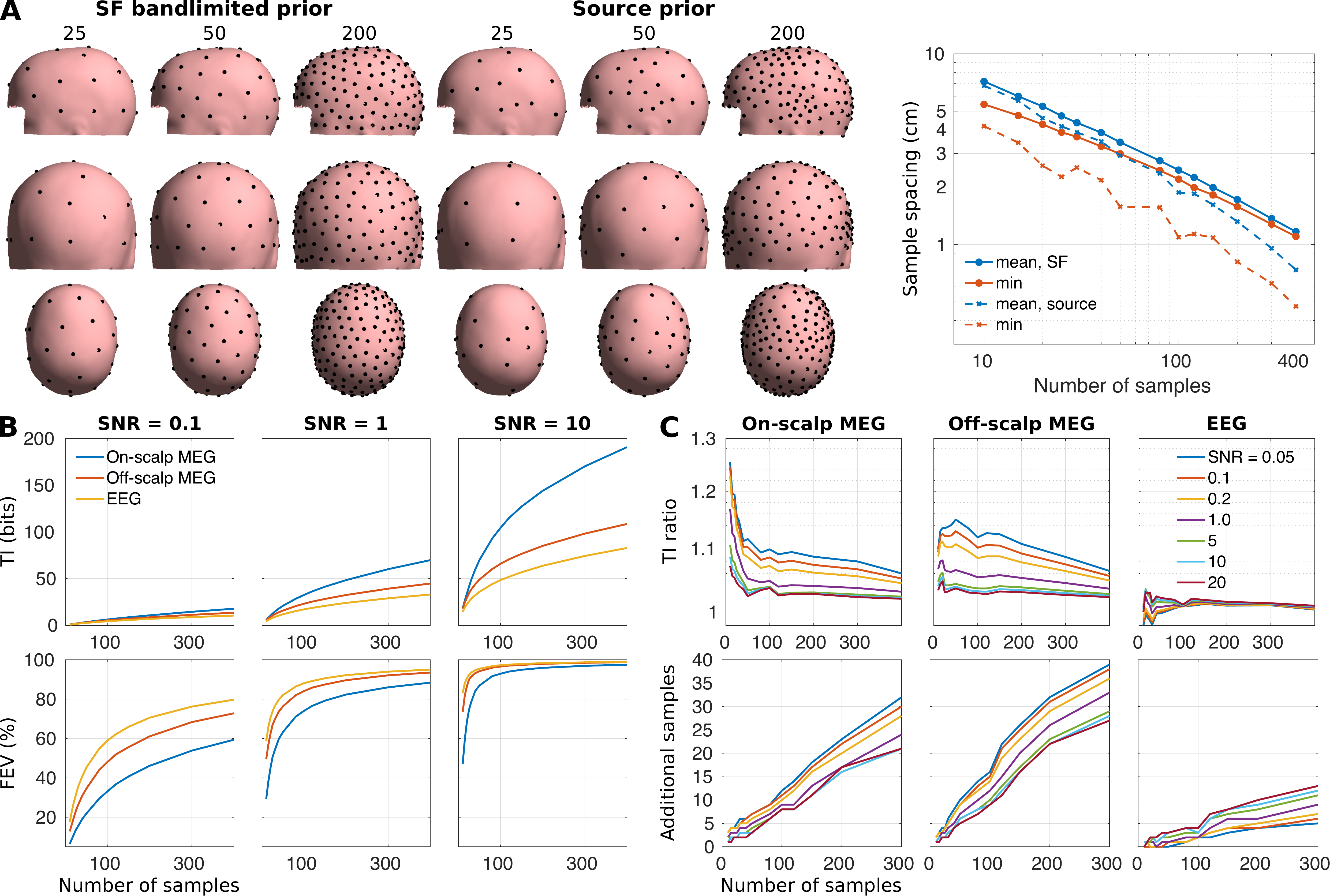}
\caption{Comparison of uniform (SF bandlimited prior) and model-informed (IID source prior) sampling of the whole brain in on-scalp MEG, off-scalp MEG and EEG. {\bf A:} Example on-scalp MEG sampling grids generated with the priors and their average and minimum sample spacing. {\bf B:} Total information (TI) and fractional explained variance (FEV) computed with different SNRs for uniform grids. {\bf C:} Ratio of the TI obtained by model-informed and uniform sampling with different SNRs (top) and the number of additional samples required with uniform sampling to achieve the same TI as with model-informed sampling (bottom). }
\label{fig:iidsampling}
\end{figure*}

\paragraph{Results} Fig. \ref{fig:iidsampling} summarizes the first analysis in the global scenario. For model-informed sampling, the sample density is the highest on the temporal lobe, which corresponds to the shorter correlation lengths and higher variance of the topography kernel shown in previous analysis (Fig.~\ref{fig:fieldcovariance}). Model-informed sampling is especially beneficial in MEG at low SNR $<$ 1 and low sample numbers, giving about 10--25$\%$ increase in TI compared to uniform sampling. With model-informed sampling, the same TI can be obtained with fewer samples than with uniform sampling  (Fig. \ref{fig:iidsampling}C). The benefit from model-informed sampling decreases as the SNR or sample number increases in MEG. Further, as the number of samples increases, the benefit of each sample decreases.

With a similar SNR, the TI for a given number of samples is highest in on-scalp MEG. When the SNR is low (0.1), 60, 74 and 80$\%$ of variance is explained (on-scalp MEG, off-scalp MEG and EEG, respectively) with a sample number as high as 400. When the SNR is high (10), about 80, 30 and 20 samples are needed to explain 90$\%$ variance.

\begin{figure*}[!t]
\centering
\includegraphics[width=1.0\linewidth]{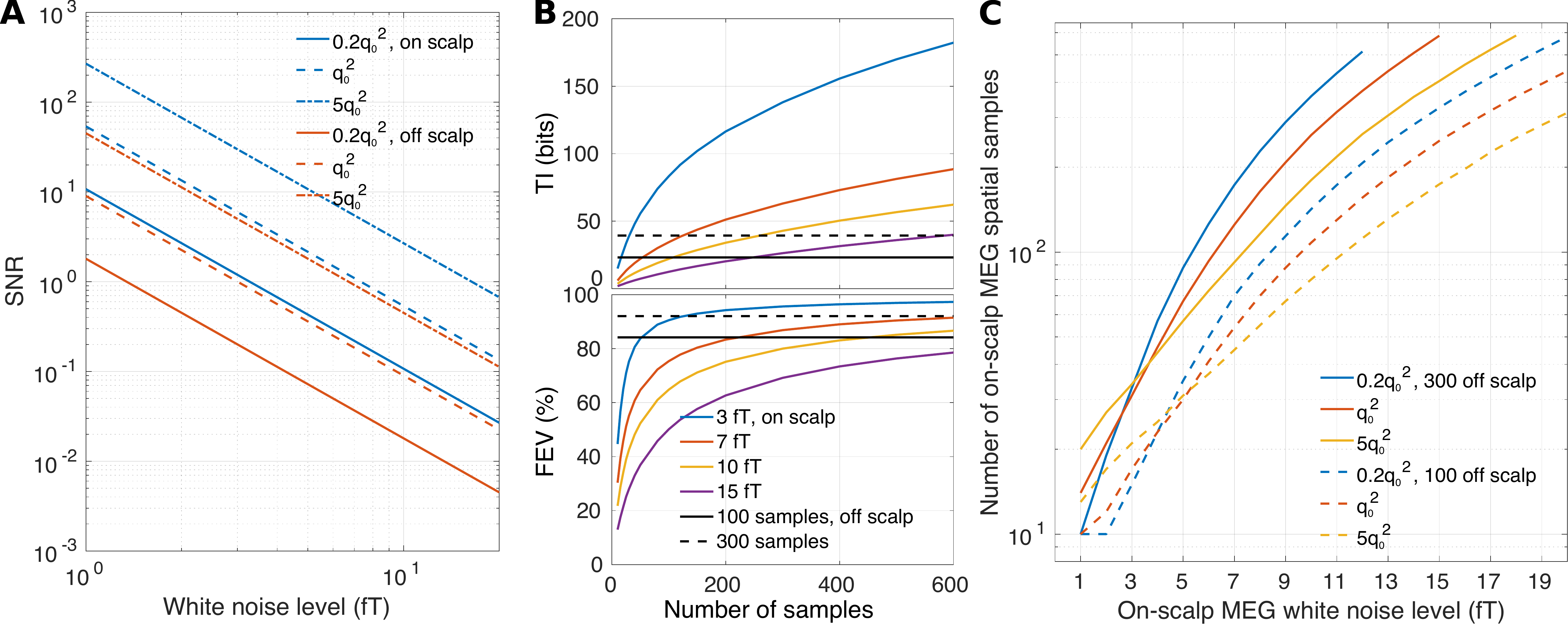}
\caption{Spatial sampling of independent sources (variance $q_0^2$) distributed over the whole cortex with on-scalp and off-scalp MEG. {\bf A:} Signal-to-noise ratio averaged over the whole measurement surface as a function of spatial white noise level. {\bf B:} Total information (TI) and fractional explained variance (FEV). Colored lines give the values for on-scalp MEG as a function of sample number and noise level. The vertical lines show the values for off-scalp MEG with 100 and 300 samples with a noise level of 3 fT. The source variance is $q_0^2$. {\bf C:} Number of spatial samples needed in on-scalp MEG to achieve the same TI as in off-scalp MEG (noise level 3 fT; 100 or 300 samples) as a function of on-scalp MEG white noise level.}
\label{fig:onscalpoffscalp}
\end{figure*}

Fig. \ref{fig:onscalpoffscalp} gives a summary of the second analysis comparing on- and off-scalp MEG. With a similar noise level, SNR as averaged over the whole measurement surface is higher in on-scalp MEG by a factor of 6.0. Fig. \ref{fig:onscalpoffscalp}B shows TI and FEV as a function number of spatial samples and noise. With a comparable noise level, on-scalp MEG achieves the same information as 300 off-scalp samples with fewer samples.

Fig. \ref{fig:onscalpoffscalp}C gives the number of spatial samples needed in on-scalp MEG for the same TI as in off-scalp MEG as a function of noise level. For example, when the source variance equals $q_0^2$, 31, 125 and 260 samples are needed in on-scalp MEG with noise levels 3, 7 and 10 fT, respectively, to yield the same TI as 300 off-scalp MEG samples. The spacings corresponding to those sample numbers in on-scalp MEG are about 4.4, 2.2 and 1.5 cm (Fig. \ref{fig:iidsampling}A). For a noise level of 15 fT, roughly 590 samples are needed. For a lower source variance $0.2q_0^2$, corresponding sample numbers for equal TI are 33, 173 and 357, while for higher variance $5q_0^2$ they are 34, 92 and 180. With a higher source variance, about 400 on-scalp samples with a noise level of 15 fT give the same TI as 300 off-scalp samples.

\subsubsection{Local scenario}

\begin{figure*}[!t]
\centering
\includegraphics[width=1.0\linewidth]{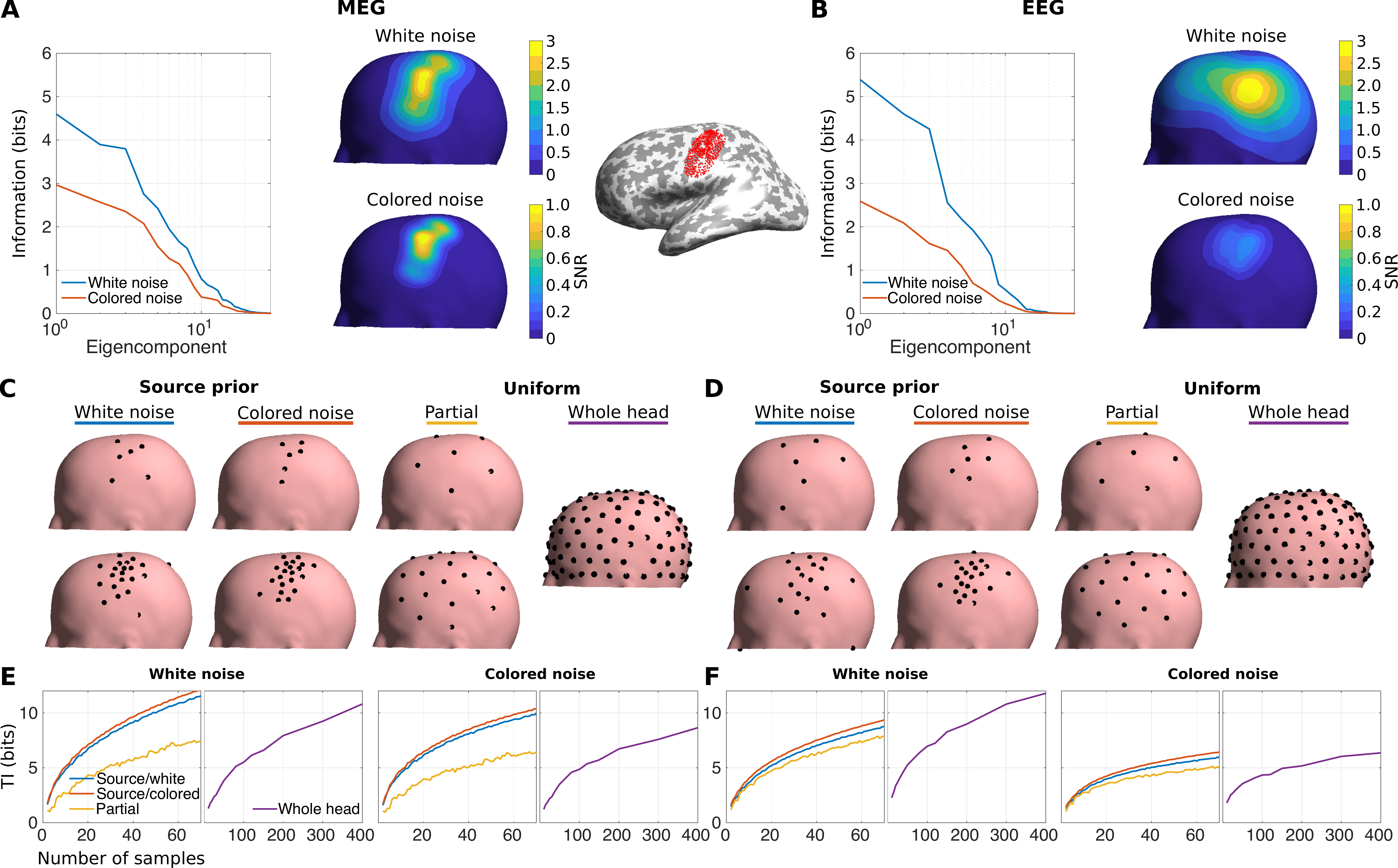}
\caption{Spatial sampling of on-scalp MEG (left panel; A, C and E) and EEG (right; B, D and F) topographies due to uncorrelated sources in an ROI around the motor cortex (shown on the center in red on the inflated cortical surface). {\bf A \& B:} Eigencomponent distribution of total information (TI) and spatial distribution of SNR computed with spatial white noise as well as with colored noise (brain background activity + white noise). {\bf C \& D:} Example grids constructed with IID source and bandlimited SF ($\sim$uniform sampling) priors. With IID source prior, two noise priors (white and colored) were considered. {\bf E \& F:} TI as a function of number of samples in the grids computed using the two noise models. }
\label{fig:roisampling}
\end{figure*}

\paragraph{Methods} In the local scenario, the ground-truth model consisted of IID sources distributed in an ROI defined around the motor cortex. Both white (sensor) and colored (sensor + background brain activity) noise were considered. The ROI consisted of sources within a patch that had a radius of approximately 3 cm along the cortical surface (387 source elements; Fig. \ref{fig:roisampling}A). Sampling grids were constructed for on-scalp MEG and EEG.

The source variance $q^2$ was set so that the maximum SNR of the on-scalp MEG was 3 with a spatial white noise level of 9 $\mathrm{fT}$. The white noise level of EEG (24 $\mathrm{nV}$) was set so that the maximum SNR was also 3. To model colored noise due to brain background activity, the sources outside the ROI were assumed active (IID Gaussian; variance $q^2/100$).

The topography kernel was whitened as described in Sec. \ref{sec:bioemsampling} and SNR was analyzed. The eigenvalues $\lambda_i$ of the whitened kernel were converted to bits as $1/2 \log_2 (\lambda_i + 1)$. These values were used to analyze the number of eigencomponents contributing to the total information. Two uniform sampling schemes were considered. In the first, samples were distributed evenly on the whole head. In the second, the SF basis was constrained to a region on the surface that had a distance to the center of the ROI less than 9 cm. For model-informed sampling, a kernel generated with IID random sources in the ROI was used, while the noise kernel corresponded either to the white or colored noise model.

\paragraph{Results} Fig. \ref{fig:roisampling}A shows the SNR distributions as well as the TI content across the eigencomponents. With white noise, the TI is 28 bits in MEG and 27 bits in EEG. In MEG, 19 eigencomponents have a TI content larger than 0.1 bits, while this number is 13 in EEG. With colored noise, the maximum SNR is 1.1 in MEG and 0.3 in EEG while the TI is 18 and 12 bits, respectively. The number of eigencomponents above 0.1 bits is in this case 16 and 12 in MEG and EEG, respectively.

The grids constructed using the different priors for signal and noise are presented in Fig. \ref{fig:roisampling}B. Compared to uniform sampling, the model-informed grids are more densely distributed, especially when the colored noise model is used. Compared to whole-head sampling, distributing samples on the region with high SNR (Fig. \ref{fig:roisampling}A) is beneficial in terms of TI both in EEG and MEG. Grids constructed with source priors give the highest TI, especially in MEG (Fig. \ref{fig:roisampling}C). For example, to reach 5 bits of information in the situation with white noise, 11 samples are needed in the model-informed MEG grids while 80 samples are needed in the whole-head grid; for EEG the corresponding sample counts are 19 and 50. With colored noise, corresponding sample counts are about 14 and 120 in MEG and 40 and 200 in EEG, respectively.

\section{Discussion}

\noindent We quantified the number of spatial samples beneficial for MEG and EEG by analyzing the spatial frequency content of the source topographies. Based on a review of Gaussian processes and experimental design, we related the information metric used in MEG sensor-array designs to the covariance of random-field models. This relationship was used to derive an information-maximizing sampling algorithm. We applied the algorithm to generate sampling grids, which we used to quantify the benefit of model-informed sampling of random source distributions in comparison to uniform sampling.

\subsection{Topography analysis}

\noindent To capture 99\% of the topography energy of every source in the brain, approximately 300, 100 and 110 SF components were needed in on-scalp, off-scalp MEG and EEG, respectively. These numbers represent the maximum number of spatial degrees of freedom in any topography due to neural activity. Additionally, they correspond to the number of uniform samples needed to achieve at most roughly 1\% reconstruction error in noiseless conditions. These results are in good accordance with previously published results. For MEG, this corresponds to the "rule of thumb" presented by Ahonen et al. \citeyear{ahonen1993}: the sensor spacing should be approximately the distance of the sensors to the closest sources. For EEG, this is in line with the arguments by Srinivasan et al. \citeyear{srinivasan1998}: the sensor count should be at least roughly 120 for adequate sampling of the EEG in a typical adult head.

The spatial bandwidths of source topographies did not depend on the level of detail of the volume conductor in MEG. In contrast, the bandwidths in EEG were affected by conductivity differences within the head, demonstrating the well-known spatial low-pass filtering of EEG \cite{srinivasan1996}. Therefore, the beneficial sensor spacing and spatial correlations in MEG depend mainly on the measurement distance while those in EEG have a dependence on the tissue conductivity contrasts.

The eigenbasis of the topography kernel of IID sources was shown to reduce the component number needed to capture 99\% of the source topography energies, compressing the topography representations on average by 19, 33 and 27\% compared to the SF basis (on-scalp, off-scalp MEG, and EEG). This compression and the variance analysis of IID sources suggest that the sample numbers given by the SF analysis provide oversampling, and, thereby, give a good upper limit on the beneficial sample number.

\subsection{Grid construction}

\noindent To generate sampling grids, we presented a method that utilizes prior information of signal and noise in the form of a covariance kernel. Assuming random spatial-frequency coefficients up to some bandlimit, the kernel is isotropic and translation-invariant and the method yields uniform sampling grids similar to those in the sampling theorems introduced in Sec. \ref{sec:sfbasis}. With this approach, uniform grids may be constructed also on more complex surfaces and domains for applications that require uniform sampling. The kernel constructed from the topographies of IID random neural sources has a location-dependent shape and length, and the method yields nonuniform sampling grids.

In the case of whole-head sampling of MEG topographies due to random sources distributed over the cortex, nonuniform model-informed sampling was only slightly beneficial compared to uniform sampling. When the number of samples was small and the noise level high, nonuniform sampling yielded roughly 10--25\% more information than uniform sampling. When decreasing the noise level or increasing the number of samples, the performance of uniform and nonuniform sampling grids became roughly similar. With nonuniform sampling, the same total information was, however, reached with fewer number of samples; the sample number was proportional to the total sample count and did not strongly depend on the SNR. In EEG, the benefits of nonuniform sampling were less clear; EEG may not benefit from it due to long-range correlations. Though, the nonuniform grids generated with the method may have been suboptimal for EEG: the method assumed positive correlations while EEG has substantial negative correlations at long distances (Fig. \ref{fig:fieldcovariance}). Altogether, among the modalities, with similar SNR and the number of samples, on-scalp MEG yielded the most information, and more measurements were needed to explain the same amount of variance.

When a cortical region of interest was defined, local high-density (nonuniform) spatial sampling was beneficial; dense sampling of the local components yielded higher total information than uniform sampling covering a larger surface area or the whole head. Dense sampling may be especially useful when colored noise fields (such as those generated by background brain activity) are present and the region of interest is small. Local dense arrays could be beneficial in applications where a certain part of the brain is of interest and the number of sensors is limited (e.g., \citealt{iivanainen2019potential}) or brain--computer interfacing where simple measurement setups are desirable.

\subsection{Future directions}

\noindent The sampling grid optimization presented here does not necessarily result in sensor arrays that can be readily implemented in practice. For example, the minimum distance between the sensors was not constrained and the sensor dimensions were ignored. Modeling sensor dimensions is important as they limit the minimum distance between the sensors and the sensor noise level is proportional to them in many sensor types (e.g., \citealt{mitchell2019quantum, kemppainen1989}). Moreover, spatial integration within a sensor can be viewed as spatial low-pass filtering, an effect that can be used to reduce aliasing due to high spatial frequencies \cite{roth1989using}. 

The optimization method at its current stage is useful for understanding how to place sensors in specific experiments such as when a certain brain region is measured. It can also be used to aid in practical sensor-array design to inform whether it is beneficial to decrease sensor noise, add sensors or switch from uniform to nonuniform sampling. In the future, the sensor dimensions and a constraint on smallest allowed sensor distance could be included, and the method could be used with a population of head models so that a sensor array with the best average performance could be designed.

We analyzed scalar fields which is sufficient in EEG. In contrast, magnetic field is a vector field, and the orthogonal field components may each provide additional information \cite{iivanainen_measuring_2017}. The sensor orientation could be included in the optimization by applying vector basis functions to assemble the kernel. Also (external) interference fields could be taken into account. In this case, the vector-spherical harmonics could be useful \cite{taulu2005presentation}; sampling grids with multiple layers of vector samples at different distances from the scalp may be beneficial \cite{nurminen2010improving, nurminen2013improving}.

With the generalization of the SF basis to an arbitrary measurement surface, the standard 1-D signal processing methods such as filtering and interpolation should be straightforward to apply to spatial EEG/MEG data. Spatial-frequency filtering could be useful in noise and interference rejection as demonstrated by Graichen et al. (\citeyear{graichen2015sphara}); the continuous SF basis functions make the filtering less dependent on the sensor configuration. As the SF basis is both data- and model-independent, the filtering methods may be useful in real-time applications, such as brain--computer interfaces. The SF filtering as a part of preprocessing would be similar to the spline Laplacian \cite{nunez2019multi}, providing flexibility for defining the filter coefficients. Filtering and interpolation may also help in data visualization.

Estimating the neural sources that could have generated the data, i.e., solving the bioelectromagnetic inverse problem \cite{sarvas1987basic} is often of interest. We did not explicitly consider the inverse problem. However, the total information should be an estimator of the source-estimation performance as it quantifies the size (or the stability of the inversion) of the modelled or measured measurement covariance matrix. In other terms, when the kernel is generated using source topographies, total information maximization corresponds to minimization of the posterior entropy of the sources (Sec. \ref{sec:optimalcriteria}). Moreover, the Gaussian random-field model has a direct connection to minimum-norm source estimation (\ref{sec:appendix_mne}).

\section{Conclusions}

\noindent We analyzed the spatial sampling of EEG and MEG and suggested a method for designing optimal sampling positions. Our simulations of neuronal fields suggest that on-scalp MEG can benefit from 300 spatial samples while for off-scalp MEG and EEG the beneficial sample number is roughly 100. Our method for sample positioning can be used to design sampling grids that convey the most information from the neuronal sources. Such designs may be useful when the sensor number is limited or a certain region of the brain is of interest.

\section*{Conflicts of interest}
\noindent The authors declare that they have no conflict of interest.

\section*{Data and code availability statement}
\noindent To facilitate further use in the research community, the information-maximizing sampling algorithm is available from the corresponding authors for academic use.

The data and code sharing policy adopted by the authors complies with the requirements of Aalto University.

\section*{Acknowledgments}
\noindent This work has received funding from the European Union’s Horizon 2020 research and innovation programme under grant agreement No 686865 (project BREAKBEN), the European Research Council under grant agreement No 678578 (project HRMEG), the National Institute of Neurological Disorders and Stroke of the National Institutes of Health under Award Number R01NS094604, the Finnish Cultural Foundation under grant nos. 00170330 and 00180388 (JI), and Vilho, Yrjö and Kalle Väisälä Foundation (AM). The content is solely the responsibility of the authors and does not necessarily represent the official views of the funding organizations.

\bibliographystyle{model5-names}
\bibliography{refs.bib}

\appendix
\section{Interpretations of posterior variables}
\label{sec:appendix_mne}
\noindent The covariance kernel obeys the form $K(\vec{r}, \vec{r}\,') = \boldsymbol{\psi}(\vec r)^\top\mathbf{K}_a\boldsymbol{\psi}(\vec r')$. As $\mathbf{k}(\vec{r}) = \boldsymbol{\Psi}\mathbf{K}_a\boldsymbol{\psi}(\vec{r})$, we can read from Eq.~\eqref{eq:posterior_bayes} that the posterior covariance of the coefficients $\mathbf{a}$ is
\begin{equation}
    \mathbf{K}_a^* = \mathbf{K}_a - \mathbf{K}_a\boldsymbol{\Psi}^\top
    (\mathbf{K} + \boldsymbol{\Sigma})^{-1}
    \boldsymbol{\Psi} \mathbf{K}_a.
\end{equation}
Application of Woodbury's matrix identity \cite{petersen2008matrix} yields
\begin{equation}
\begin{split}
    \mathbf{K}_a^*
        &= \mathbf{K}_a - \mathbf{K}_a\boldsymbol{\Psi}^\top
    (\mathbf{K} + \boldsymbol{\Sigma})^{-1}
    \boldsymbol{\Psi} \mathbf{K}_a
    \\ &= \mathbf{K}_a - \mathbf{K}_a\boldsymbol{\Psi}^\top
    (\boldsymbol{\Psi} \mathbf{K}_a \boldsymbol{\Psi^\top} + \boldsymbol{\Sigma})^{-1}
    \boldsymbol{\Psi} \mathbf{K}_a
    \\ &=(\boldsymbol{\Psi}^\top \boldsymbol{\Sigma}^{-1} \boldsymbol{\Psi} + \mathbf{K}_a^{-1})^{-1},
\end{split}
\end{equation}
which approaches the covariance of the coefficient error of a bandlimited topography (Eq.~\eqref{eq:fieldrecoerr}), when $\mathbf{K}_a^{-1}$ approaches zero and the measurement noise is white. 

Similarly, the posterior mean of the coefficients is 
\begin{equation}
    \boldsymbol{\mu}_a^* = \mathbf{K}_a\boldsymbol{\Psi}^\top (\mathbf{K} + \boldsymbol{\Sigma})^{-1}\mathbf{y} 
    =\mathbf{K}_a\boldsymbol{\Psi}^\top (\boldsymbol{\Psi \mathbf{K}_a \Psi^\top} + \boldsymbol{\Sigma})^{-1}\mathbf{y},
\end{equation}
which follows the form of the generalized minimum-norm estimate \cite{dale1993improved}. Inserting $\boldsymbol{\Sigma} = \sigma^2\mathbf{I}$ and $\mathbf{K}_a = \lambda^2\mathbf{I}$, we get the coefficients to the form $\boldsymbol{\Psi}^\top(\boldsymbol{\Psi}\boldsymbol{\Psi}^\top  + (\sigma^2/\lambda^2)\mathbf{I})^{-1}\mathbf{y}$, which is the more common form for the estimator, $\sigma^2/\lambda^2$ being the Tikhonov regularization parameter.

The coefficient estimator can be further manipulated:
\begin{equation}
    \begin{split}
    \boldsymbol{\mu}_a^* &= \mathbf{K}_a\boldsymbol{\Psi}^\top (\mathbf{K} + \boldsymbol{\Sigma})^{-1}\mathbf{y} = \mathbf{K}_a\boldsymbol{\Psi}^\top (\mathbf{K} + \boldsymbol{\Sigma})^{-1}[(\mathbf{K} + \boldsymbol{\Sigma}) - \mathbf{K}] \boldsymbol{\Sigma}^{-1}\mathbf{y} \\
    &= \mathbf{K}_a\boldsymbol{\Psi}^\top [\mathbf{I} - (\mathbf{K} + \boldsymbol{\Sigma})^{-1}\mathbf{K}] \boldsymbol{\Sigma}^{-1}\mathbf{y} \\
    &= [\mathbf{K}_a\boldsymbol{\Psi}^\top - \mathbf{K}_a\boldsymbol{\Psi}^\top(\mathbf{K} + \boldsymbol{\Sigma})^{-1}\boldsymbol{\Psi}\mathbf{K}_a\boldsymbol{\Psi}^\top] \boldsymbol{\Sigma}^{-1}\mathbf{y} \\
    &= [\mathbf{K}_a - \mathbf{K}_a\boldsymbol{\Psi}^\top(\mathbf{K} + \boldsymbol{\Sigma})^{-1}\boldsymbol{\Psi}\mathbf{K}_a] \boldsymbol{\Psi}^\top\boldsymbol{\Sigma}^{-1}\mathbf{y} = \mathbf{K}_a^*\boldsymbol{\Psi}^\top\boldsymbol{\Sigma}^{-1}\mathbf{y} \\
    &= (\boldsymbol{\Psi}^\top \boldsymbol{\Sigma}^{-1} \boldsymbol{\Psi} + \mathbf{K}_a^{-1})^{-1}\boldsymbol{\Psi}^\top\boldsymbol{\Sigma}^{-1}\mathbf{y},
    \end{split}
\end{equation}
which again reduces to the least-squares estimator of Eq.~\eqref{eq:ord_least_squares}, when $\mathbf{K}_a^{-1}$ approaches zero and the measurement noise is white.

\section{Total information and covariance}
\label{sec:apppendix_info}
\noindent Channel capacity or the total information (TI) conveyed by a noisy channel \cite{shannon1949} has been used to evaluate different sensor arrays in MEG \cite{kemppainen1989, nenonen2004total,iivanainen_measuring_2017,riaz2017evaluation}. Here, we show how the channel capacity relates to the sample and noise covariance matrices, $\mathbf{K}$ and $\boldsymbol{\Sigma}$ (Sec.~\ref{sec:bayes_sampling}).

If the samples as well as the noise were uncorrelated, TI would be $1/2\sum_i \log_2(P_i + 1)$ where $P_i$ is the power signal-to-noise ratio of each measurement. If the measurements are correlated, they can be orthogonalized by the eigendecomposition of the whitened covariance matrix $\Tilde{\mathbf{K}} = \boldsymbol{\Sigma}^{-1/2}\mathbf{K}\boldsymbol{\Sigma}^{-\top/2} = \mathbf{V}\mathbf{P}\mathbf{V}^\top$, where $\mathbf{V}$ contains eigenvectors of $\Tilde{\mathbf{K}}$ and $\mathbf{P}$ is a diagonal matrix of $P_i$.
Starting from the original formula, TI can be now written as 
\begin{equation}
\begin{split}
\mathrm{TI}(R) &= \frac{1}{2}\sum_i \log_2(P_i + 1) = \frac{1}{2} \log_2 \det(\mathbf{P} + \mathbf{I}) 
\\ &=\frac{1}{2} \log_2 \det(\tilde{\mathbf{K}} + \mathbf{I}) 
= \frac{1}{2} \log_2 \frac{\det(\mathbf{K} + \boldsymbol{\Sigma})}{\det(\boldsymbol{\Sigma})}.
\end{split}
\end{equation}

The determinant measures the "size" of $\mathbf{K} + \boldsymbol{\Sigma}$. The determinants can also be interpreted as volumes spanned by the possible measurements and noise in the $N$-dimensional signal space ($N$ being the number of channels). This leads us to Shannon's original geometric interpretation: information measures (logarithmically) how many distinct signals are there in the signal space when each signal is surrounded by a volume of uncertainty $\det(\boldsymbol{\Sigma})$.

Total information can be maximized by choosing the measurement grid so that $\tilde{\mathbf{K}}$ is diagonal, i.e., each sample measures independent information. This can be seen, e.g., from the matrix derivative \cite{petersen2008matrix}
\begin{equation}
    \frac{\partial \ln\, \det(\mathbf{X} + \mathbf{I}) }{\partial \mathbf{X}} = 2(\mathbf{X} + \mathbf{I})^{-1} - \mathbf{I} \odot (\mathbf{X} + \mathbf{I})^{-1},
\end{equation}
where $\mathbf{I} \odot$ is element-wise product with identity matrix resulting in the diagonal-part of the matrix. The logarithm of determinant is maximized when the derivative with respect to its elements is zero, which means that the non-diagonal elements of $(\mathbf{X} + \mathbf{I})^{-1}$ must be zero. This is equivalent of $\mathbf{X}$ being diagonal itself. For diagonal elements, no solution exists, but their derivatives approach zero as the elements themselves approach infinity. This makes sense as TI grows logarithmically with the signal variance.

\end{document}